# Quasi-Linear Cellular Automata

Cristopher Moore
Santa Fe Institute

October 22, 2018

**Abstract**

Simulating a cellular automaton (CA) for $t$ time-steps into the future requires $t^2$ serial computation steps or $t$ parallel ones. However, certain CAs based on an Abelian group, such as addition mod 2, are termed *linear* because they obey a principle of superposition. This allows them to be predicted efficiently, in serial time $\mathcal{O}(t)$ or $\mathcal{O}(\log t)$ in parallel.

In this paper, we generalize this by looking at CAs with a variety of algebraic structures, including quasigroups, non-Abelian groups, Steiner systems, and others. We show that in many cases, an efficient algorithm exists even though these CAs are not linear in the previous sense; we term them *quasilinear*. We find examples which can be predicted in serial time proportional to $t$, $t \log t$, $t \log^2 t$ and $t^\alpha$ for $\alpha < 2$, and parallel time $\log t$, $\log t \log\log t$ and $\log^2 t$.

We also discuss what algebraic properties are required or implied by the existence of scaling relations and principles of superposition, and exhibit several novel "vector-valued" CAs.

## 1 Introduction: CAs as algebras

A *Cellular Automaton* is a dynamical system on sequences $\ldots a_{-1}a_0a_1\ldots$ where the $a_i$ are symbols in some finite alphabet $A$. It is updated by a local function $f$, usually written

$$a_i' = f(a_{i-r}, \ldots, a_i, \ldots, a_{i+r})$$

where $r$ is the *radius* of the rule.

In this paper, we will consider $r = 1/2$ rules, with a staggered space-time as shown in figure 1; a CA of any radius can be transformed into one with $r = 1/2$ by grouping sets of $2r$ sites together. Each new state is a function of just two predecessors,

$$a_i' = f(a_{i-1/2}, a_{i+1/2})$$

or

$$a = b \bullet c$$

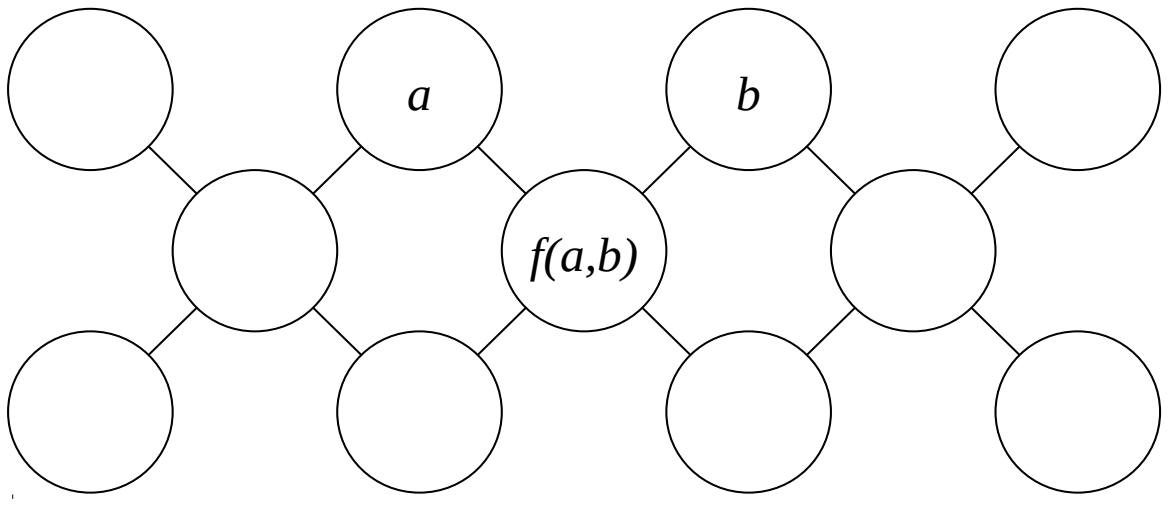

Figure 1: The staggered space-time of an $r = 1/2$ CA.

where $\bullet$ is a binary operation on $A$. This approach was also taken in [2, 3].

Like any dynamical system, we would like to know how hard a particular CA is to predict. In particular, suppose we are given a finite stretch of initial conditions, $a_0 \ldots a_t$. After $t$ time-steps, there is a single site at the bottom of a light-cone with this initial row on top. Call the state of the CA at that site $s = P_t(a_0, a_1, \ldots, a_t)$, where $P_t$ is the *prediction function*.

In general, $P_t$ is quite messy; the light-cone looks like

$$\begin{array}{ccccccc} a_0 & & a_1 & & a_2 & & a_3 \\ & a_0a_1 & & a_1a_2 & & a_2a_3 & \\ & & (a_0a_1)(a_1a_2) & & (a_1a_2)(a_2a_3) & & \\ & & & s = ((a_0a_1)(a_1a_2))((a_1a_2)(a_2a_3)) & & & \end{array} \quad (1)$$

where we have abbreviated $a \bullet b$ as simply $ab$. Clearly, $P_t$ can always be calculated in $t(t+1)/2 = \mathcal{O}(t^2)$ serial steps or $t$ parallel ones, simply by simulating the CA and calculating all the products in the light-cone above $s$. But if $\bullet$ fulfills certain algebraic properties, we may be able to calculate $P_t$ much more efficiently. Discussing how different algebraic structures on $\bullet$ affects the complexity of $P_t$ is the aim of this paper.

There is an enormous amount of variety between easily predictable CAs, such as addition mod 2, and computationally universal ones, such as in [4]. We hope to begin to map out this region, rising a few steps above the very simplest rules, and explore the grey area between linearity and computational universality.

This may be relevant to the use of CAs in cryptography; several partially linear CA rules have been proposed as cryptosystems, either as random number generators [5] or as iterated encrypting functions [6]. Obviously these CAs would be much less useful for this purpose if they turn out to be efficiently predictable.

## 2 Preliminaries

We say $(A, \bullet)$ is a *quasigroup* if the *left and right division properties* hold: i.e., for all $a, b$ there is some $c$ such that $a \bullet c = b$, and some $d$ such that $d \bullet a = b$. We can also express this by saying that left and right multiplication by any element $a$ is a one-to-one operation. If an $r = 1/2$ CA is a quasigroup, it is one-to-one on both its inputs; this is also referred to as *left and right permutive*.

If a CA is a quasigroup, this ensures that every row has exactly $n = |A|$ pre-images, since we can choose one site in the pre-image and then use left and right division to fill in the rest of the sites. So quasigroup CAs are *surjective* (onto) but not reversible (since there's more than one pre-image).

The multiplication tables of quasigroups are *Latin squares*, $|A|$ by $|A|$ squares where each element occurs exactly once in each row and each column [12].

A *left (right) identity* is an element $e$ such that $e \bullet a = a$ ($a \bullet e = a$). An *identity* is both a left and a right identity. A quasigroup with identity is called a *loop*. In a CA, an identity will be invaded by any other state.

A *left (right) zero* is an element $z$ such that $z \bullet a = z$ ($a \bullet z = z$), so that it spreads and invades all other states. Clearly no quasigroup can have a zero.

An element $a$ is *idempotent* if $a \bullet a = a$. Domains of idempotents will persist, possibly invaded from left and right, so they will typically form triangular domains in the space-time history of the CA.

We say $\bullet$ is *associative* if $(a \bullet b) \bullet c = a \bullet (b \bullet c)$ for all $a, b, c$. $(A, \bullet)$ is then a *semigroup*.

If an identity exists, a *left (right) inverse of* $a$ is an element $a^{-1}$ such that $a^{-1} \bullet a = e$ ($a \bullet a^{-1} = e$). Left and right inverses are the same if $\bullet$ is associative.

If $\bullet$ is associative, has an identity $e$, and has an inverse $a^{-1}$ for every $a$, then $(A, \bullet)$ is a *group*. Every associative quasigroup is a group, since the division properties allow us to find inverses and an identity.

We say $\bullet$ is *commutative* if $a \bullet b = b \bullet a$ for all $a, b$. Commutative groups are also called *Abelian*. We will often use the symbols $+$ and $0$ for $\bullet$ and $e$ respectively for an Abelian group.

The Abelian group $\{0, 1, \ldots, p-1\}$ with addition mod $p$ is called the *cyclic group* $Z_p$.

If $(A, \bullet)$ and $(B, \bullet)$ are two groups (or quasigroups, or whatever), their *direct sum* $A \oplus B$ is the set of vectors $(A \times B, \bullet)$, where $(a, b) \bullet (c, d) = (a \bullet c, b \bullet d)$.

A function $h$ from $(A, \bullet)$ to $(B, *)$ is a *homomorphism* if $h(a \bullet b) = h(a) * h(b)$. A homomorphism that is one-to-one and onto is an *isomorphism*, and we say $A$ and $B$ are *isomorphic* if one exists; i.e., $A$ and $B$ are essentially identical except for a relabelling of their elements. An isomorphism from $A$ to itself is called an *automorphism*.

Since composition $\circ$ is an associative operation, the set of functions or homomorphisms from $A$ to $A$ form a semigroup, with the identity function $1(a) = a$. Since automorphisms have inverses, they form a group.

Homomorphisms on a commutative group $(A, +)$ can also be summed, by defining $(f + g)(a) = g(a) + f(a)$; so they also form a group under $+$. The two operations $\circ$ and $+$ obey a *distributive* property, namely $f \circ (g + h) = f \circ g + f \circ h$. The map $0(a) = 0$, where $0$ is the identity of $A$, is the identity of $+$ but a zero of $\circ$. (This kind of two-operation structure is called a *ring*.) In the text, we will usually abbreviate $f \circ g$ as $fg$.

Our model of serial computation will be a *Random Access Machine* (RAM). It has an arbitrary number of registers, each of bounded length, which it can access in constant time. In each computation step it can only manipulate a finite number of bits (so, for instance, it takes $\mathcal{O}(n)$ steps to add two $n$-digit numbers). Many of our algorithms will work just as well on a multitape Turing machine [15].

Our model of parallel computation will be families of Boolean circuits. The class $\mathbf{NC}_j$ is the set of functions that can be calculated by circuits with a polynomial number of gates and depth $\log^j n$ where $n$ is the number of bits of input ($n = c(t+1)$ for our purposes, for an alphabet with $2^c$ symbols). This is equivalent to a *Parallel Random Access Machine* (PRAM) that computes the function in time $\mathcal{O}(\log^j t)$. The union of all the $\mathbf{NC}_j$ is $\mathbf{NC}$; i.e., functions that can be computed in polylogarithmic parallel time by circuits of polynomial size.

(Technical note: we are using CREW (Concurrent Read, Exclusive Write) PRAMs where several processors can read, but not write to, the same register at once. In [16] RAMs and PRAMs are defined with each register holding an arbitrary integer, and with addition as an elementary operation that can be carried out in one step. We feel that bounding the number of bits in each register gives results more relevant to real computers.

Finally, by $\log^j n$ we mean $\log n$ raised to the $j$th power, not log composed $j$ times.)

# 3 Abelian groups

If $(A, \bullet)$ is an Abelian group, we'll write $+$ and 0 instead of $\bullet$ and $e$ respectively. Equation (1) then becomes

$$\begin{array}{ccccccc} a_0 & & a_1 & & a_2 & & a_3 \\ & a_0+a_1 & & a_1+a_2 & & a_2+a_3 & \\ & & a_0+2a_1+a_2 & & a_1+2a_2+a_3 & & \\ & & & a_0+3a_1+3a_2+a_3 & & & \end{array} \tag{2}$$

So we can write

$$P_t(a_0, a_1, \ldots, a_t) = \sum_{x=0}^{t} \binom{t}{x} a_x$$

or

$$P_t = \sum_{x=0}^{t} G_{x,t}(a_x)$$

where $G$ is a *Green's function*,

$$G_{x,t}(a) = P_t(\ldots, 0, a, 0, 0, \ldots) = \binom{t}{x} a \tag{3}$$

where $a_x = a$ and all other $a_i = 0$. In this case $G_{x,t}$ is a scalar multiplication, so we will simply identify it with the Pascal's Triangle coefficient and write $G_{x,t} = \begin{pmatrix} t \\ x \end{pmatrix}$.

By the Chinese Remainder theorem, any Abelian group can be written as a direct sum $A = Z_{{p_1}^{r_1}} \oplus Z_{{p_2}^{r_1}} \oplus \cdots \oplus Z_{{p_k}^{r_k}}$ where the $p_i$ are primes — so all we need to know [1, 2] is $G_{x,t} \bmod {p_i}^{r_i}$ for each $i$.

But each of these coefficients can be calculated in the following way: since $\begin{pmatrix} t \\ x+1 \end{pmatrix} = \begin{pmatrix} t \\ x \end{pmatrix} (t - x \,/\, x + 1)$, we start with $G_{0,t} = 1$ and then use this recurrence to go from $G_{x,t}$ to $G_{x+1,t}$. We calculate $G_{x,t}$ as follows: write $G_{x,t} = (p^r)^m b$ where $p^r$ does not divide $b$. Then

$$G \bmod p^r = \begin{cases} 0 \text{ if } m \neq 0 \\ b \bmod p^r \text{ if } m = 0. \end{cases}$$

So at each step, all we need to do is change the value of $b \bmod p^r$ and increment or decrement $m$. But we can similarly write $t - x = (p^r)^{m_1} b_1$ and $x + 1 = (p^r)^{m_2} b_2$, and then

$$G_{x+1,t} = (p^r)^{m'} b' = G_{x,t} \frac{t-x}{x+1} = (p^r)^{m+m_1-m_2} b b_1 / b_2$$

If $r = 1$, then $m' = m + m_1 - m_2$ and $b' = bb_1/b_2 \bmod p$. If $r > 1$, we increment $m'$ if $bb_1/b_2$ has $r$ factors of $p$.

So as $x$ ranges from 0 to $t$, we have to:

1) calculate $m_1, m_2, b_1$ and $b_2$

2) update $m$ and $b$ accordingly; and

3) add all the $G_{x,t} a_x$ to get $s$.

On a standard digital computer, counting all the increment and carry steps, step (1) takes time proportional to the $m$'s, step (2) takes $\log m$ time to update $m$ (for the addition) and constant time to update $b$ (since arithmetic mod $p^r$ is a finite-state process) and step (3) takes constant time. So the total time is

$$\sum_{x=0}^{t} c_1(m_1 + m_2) + c_2 \log \max(m, m_1, m_2) + c_3 < c_4 t$$

since the leading behavior comes from the first term and $\sum_{x=0}^{t} m_1 < t$ (similarly for $m_2$).

So, we've reproduced the result of [1] that for an 'additive' CA, for which $(A, \bullet)$ is an Abelian group, prediction can be done in time $\mathcal{O}(t)$ using the principle of superposition — much faster than the time $\mathcal{O}(t^2)$ full simulation takes. (This is also a lower bound, since any CA that depends on all its inputs takes linear time to read them.)

In [1], a different method was used, in which we treat the Green's function as a polynomial [7] and use the scaling properties of polynomials in a finite ring. We review this technique here since we will use it later on. Let

$$G_t(x) = G_1^t(x) = (1+x)^t$$

where $x$ represents a position operator and $G_{i,t}$ is the coefficient of $x^i$ in $G_t$. Again, we just need to calculate $G$ mod a finite number of primes and prime powers.

If $p$ is a prime, a classic theorem says that, for any polynomial $P$, $P^p(x) \equiv_p P(x^p)$. In particular,

$$G_p(x) = (1+x)^p \equiv_p 1 + x^p = G_1(x^p)$$

so that

$$P_p(a_0, a_1, \ldots, a_p) = a_0 + a_p = P_1(a_0, a_p)$$

In other words, $p$ steps of the CA are equivalent to a single step on a dilated set of initial conditions: the CA obeys a *scaling relation* with a scaling ratio of $p$. For a prime power $q = p^m$, this isn't quite true, but we still have

$$G_q(x) \equiv_p G_{p^{m-1}}(x^p) \text{ so } G_{p^n}(x) \equiv_p G_{p^{m-1}}(x^{p^{n-m+1}}) \text{ for } n \geq m-1$$

so

$$P_{p^n}(a_0, a_1, \ldots, a_{p^n}) = P_{p_{m-1}}(x_0, x_{p^{n-m+1}}, \ldots, x_{p^n})$$

and the CA scales after the first $p^{m-1}$ steps.

In either case, we can use this scaling relation to calculate the final state $s$ from just a few intermediate states. For prime $p$, jump up by $p^k d_1(t)$ where $k = \lfloor \log_p t \rfloor$ is the largest power of $p$ in $t$'s base $p$ expansion, and $d_1(t)$ is its most significant digit.

We only need to know $d_1(t)+1$ sites in this row, since $(1+x)^{p^k d_1} \equiv_p (1+x^{p^k})^{d_1}$ has only $d_1+1$ non-zero terms. Then jump up $d_2(t)p^{k-1}$ from there, and so on until we get to the top row. As is apparent from figure 2, the total number of sites we need to calculate is

$$1 + (d_1+1) + (d_1+1)(d_2+1) + \cdots = \sum_{i=0}^{k} \prod_{j=1}^{i} (d_j+1)$$

which, in the worst case where $t = p^{k+1}-1$ and $d_j = p-1$ for all $j$, is

$$\sum_{i=0}^{k} p^i = \frac{1}{p-1}\, t$$

so we get our answer in linear time.

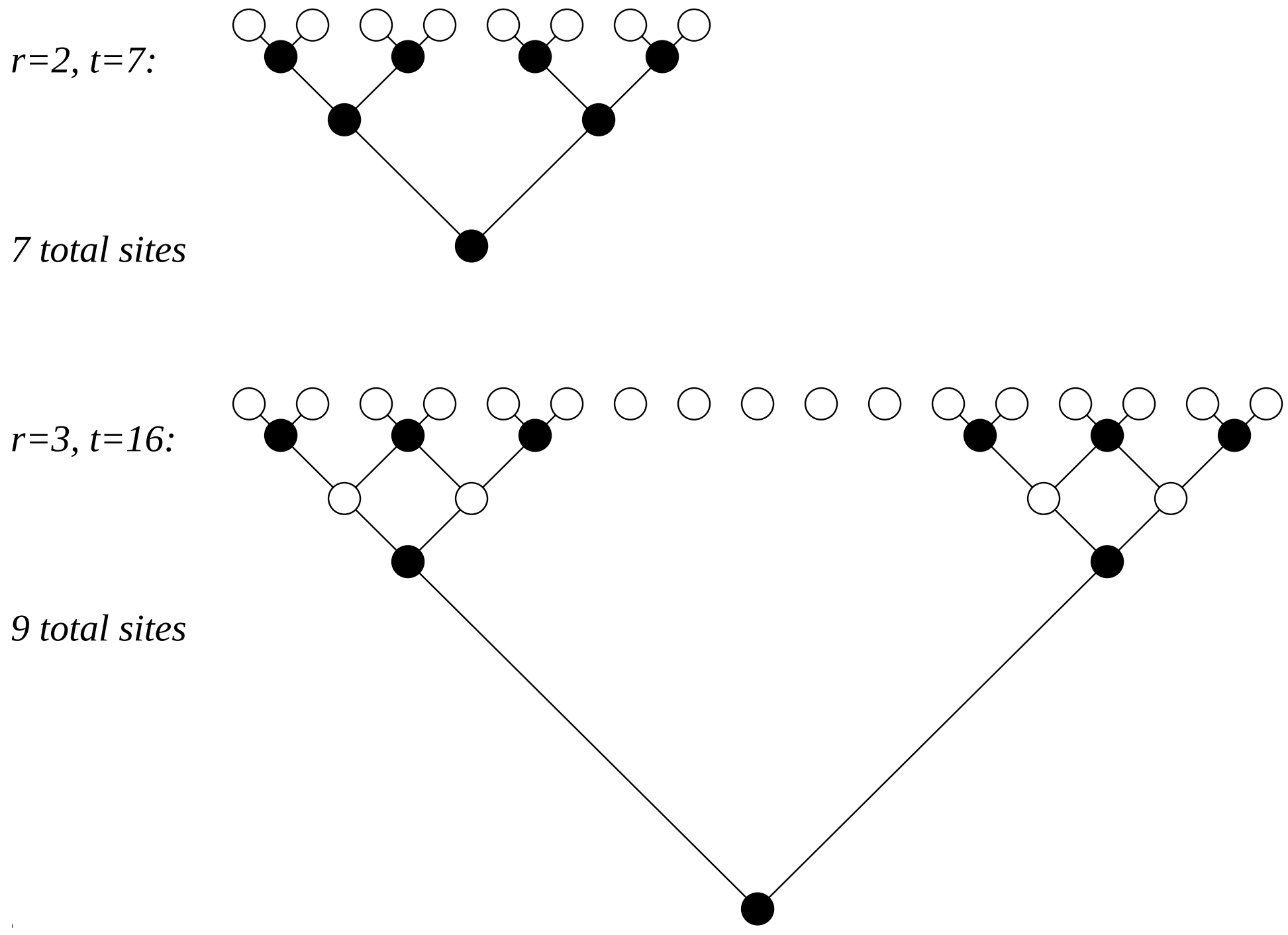

Figure 2: Using scaling for addition mod a prime $p$. The scaling ratio is $r = p$, and $(1+x)^{p^k} \equiv 1 + x^{p^k}$. The total number of sites we need in the lower case where $p = 3$ and $t = 16$ is 9 (here $d_1 = 1$, $d_2 = 2$, and $d_3 = 1$).

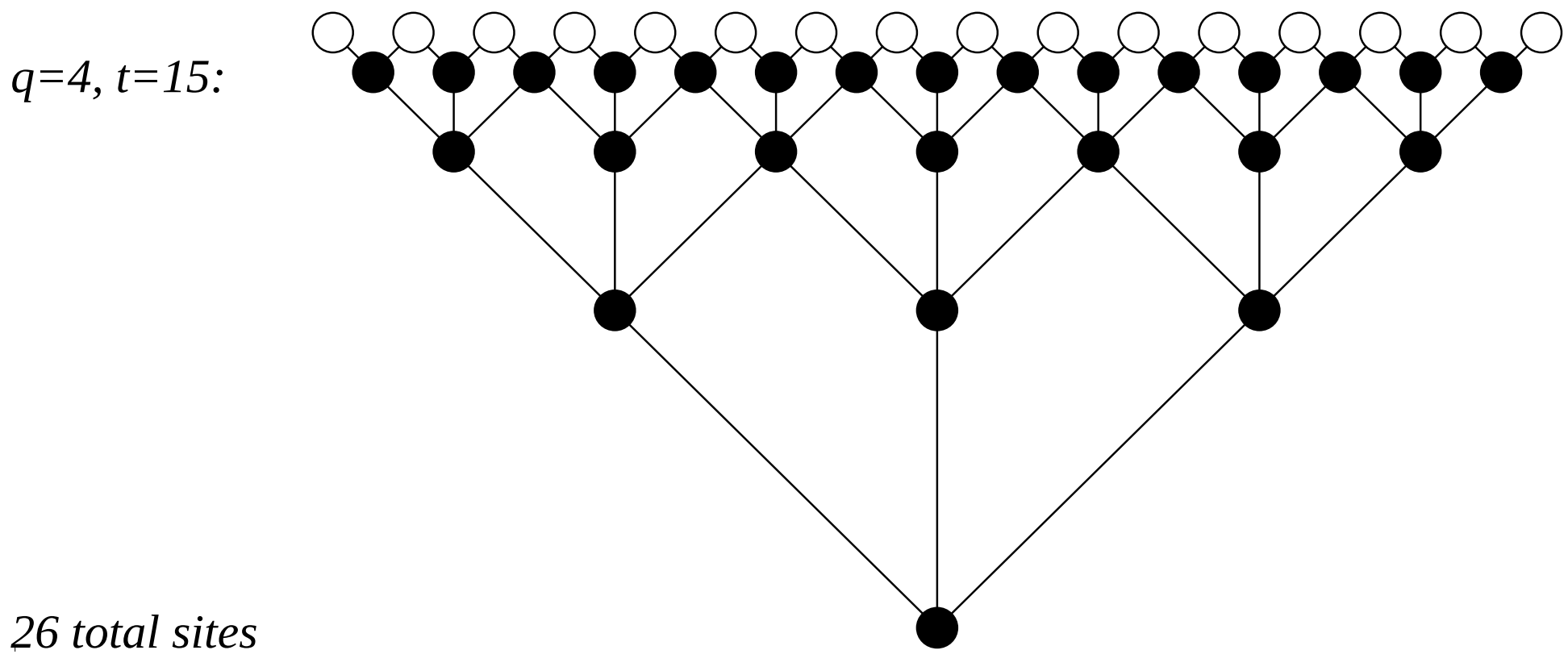

Figure 3: Scaling for addition mod a prime power $q = p^m$. Here $q = 4$, so $(1+x)^{2^k} \equiv 1 + 2x^{2^{k-1}} + x^{2^k}$. The total number of sites we need to calculate for $t = 15$ is 26.

For a prime power $q = p^m$, first simulate $t \bmod p^{m-1}$ steps (which only takes linear time) so that the remaining time is a multiple of $p^{m-1}$. Then we do the same thing, jumping up in powers of $p$, except that this time

$$(1+x)^{p^k d_1} \equiv_p (1+x^{p^{k-m+1}})^{p^{m-1}d_1}$$

has at most $p^{m-1}d_1 + 1$ non-zero terms. In addition, many of the intermediate sites we need to calculate overlap, as shown in figure 3, so the worst case is

$$\sum_{i=0}^{k-1} p^{m-1}(p^i - 1) + 1 \leq \frac{p^{m-1}}{p-1} t$$

Since we need to do this for a finite number of primes or prime powers, we have

**Theorem 1.** *An Abelian group CA can be predicted in serial time $\mathcal{O}(t)$.*

Since the scaling technique gives us a tree of depth $\log_p t$, with $\mathcal{O}(t)$ nodes, we can also say

**Theorem 2.** *An Abelian group CA can be predicted in parallel time $\mathcal{O}(\log t)$ by $\mathcal{O}(t)$ processors, so its prediction is in* $\mathbf{NC}_1$.

Now let's move beyond Abelian groups. When do the same prediction algorithms work? When do they break down, and what can replace them?

# 4 Commutative semigroups

If $\bullet$ is commutative and associative, then $(A, \bullet)$ is a commutative semigroup. Again, let's use the symbol $+$ instead of $\bullet$. If an identity doesn't already exist, we can add one, called 0, such that $a + 0 = 0 + a = a$.

Just as for Abelian groups, the light-cone (1) can be simplified into a Green's function sum (2) and (3); except that multiplication by the coefficient $G$ is actually the function $G(a) = \underbrace{a + a + \cdots + a}_{G \text{ times}}$. This is a homomorphism whenever $A$ is a commutative semigroup.

Now since $G(a) = (G-1)(a) + a$, we can start with the identity function $1(a) = a$ and add it iteratively to find what action $G(a)$ has on $A$. But since $A$ is finite, it only has a finite number of homomorphisms; so this iteration will eventually enter a finite loop. So for all $G$ greater than some $j$, $G(a)$ will only depend on $a$ and $G \bmod p$ for some $p$ — i.e., $G(a)$ is eventually periodic in $G$. So except for a number of $G$'s bounded by a constant, all we have to know is $G \bmod p$ for some $p$.

Since calculating $G \bmod p$ is equivalent to finding the Green's functions for the CA based on $Z_p$, the rest of the proof follows as above, and we have

**Theorem 3.** *A CA with a commutative semigroup operation can be predicted in time $\mathcal{O}(t)$ in serial or $\mathcal{O}(\log t)$ in parallel, so its prediction is in* $\mathbf{NC}_1$.

# 5 CAs separably isotopic to Abelian groups: Additive CAs with matrix coefficients

Two Latin squares ($n \times n$ squares where each one of $n$ symbols occurs exactly once in each row and each column) are said to be *isotopic* if they can be transformed into each other by permuting the rows, the columns, or the set of symbols [12].

Recall that Latin squares can also be thought of as the multiplication tables of quasigroups. Then we can express isotopy between two quasigroups $(A, \bullet)$ and $(A, *)$ as

$$h(a \bullet b) = g(a) * f(b)$$

where $f$, $g$, and $h$ are one-to-one functions on $A$, corresponding to permutations on the rows, columns, and symbols respectively.

Clearly, an isotopy represents a similarity in the structures of $\bullet$ and $*$, but a much looser one than an isomorphism in which $f = g = h$. In this section we will examine quasigroups that are isotopic to an Abelian group $(A, +)$, in the hope that their "group-like" structure will help us predict them efficiently.

First, if $\bullet$ is isotopic to $*$, there is always an operation $\circ$, isomorphic to $\bullet$, which is isotopic to $*$ with $h$ the identity (this is called a *principal isotope* of $*$) [12]. So it's sufficient to drop $h$ and just write

$$a \bullet b = g(a) + f(b).$$

Then, the evolution of the CA looks like

$$\begin{array}{ccccc} a_0 & & a_1 & & a_2 \\ & g(a_0) + f(a_1) & & g(a_1) + f(a_2) & \\ & \multicolumn{3}{c}{g(g(a_0) + f(a_1)) + f(g(a_1) + f(a_2))} & \end{array} \tag{4}$$

Clearly, the isotopy isn't being much help. We can only take advantage of $+$'s properties as an Abelian group if we can break terms like $g(g(a_0) + f(a_1))$ into two separate parts, one depending on $a_0$ and one depending on $a_1$.

For what $f$ and $g$ can we do that? Call a function $f$ *separable* if

$$f(a + b) = u(a) + v(b)$$

for some $u$ and $v$.

Now since $+$ is a group, it has an identity 0, and inverses $-a$ for each $a$. Then if we write $u'(a) = u(a) - u(0)$ and $v'(a) = u(0) + v(a)$, we have $f(a + b) = u'(a) + v'(b)$ with $u'(0) = 0$. So it's sufficient to assume that $u(0) = 0$.

But then $f(a) = f(0 + a) = u(0) + v(a) = v(a)$, so $f = v$, and $f(a) = f(a + 0) = u(a) + f(0)$, so $f = u + f(0)$. Moreover, $u(a + b) = f(a + b) - f(0) = u(a) + f(b) - f(0) = u(a) + u(b)$, so $u$ is a homomorphism. So we've proved that any separable function $f$ is actually a homomorphism $u$, plus a constant $f(0)$. (Note that we have not used the commutativity of $+$ to prove this, so this is true for groups in general.)

Henceforth, we will restrict our attention to quasigroups that are *separably isotopic* to an Abelian group; i.e., where $f$ and $g$ are homomorphisms plus constants, so that $f(a+b) = f(a)+f(b)-f(0)$ and $g(a+b) = g(a)+g(b)-g(0)$. (Here we are using the commutativity of $+$.) Then the light-cone (4) becomes

$$\begin{array}{ccccccc} a_0 & & a_1 & & a_2 & & a_3 \\ & g(a_0)+f(a_1) & & g(a_1)+f(a_2) & & g(a_2)+f(a_3) & \\ \multicolumn{3}{c}{g^2(a_0)+(gf+fg)(a_1)+f^2(a_2)-(f+g)(0)} & & \multicolumn{3}{c}{g^2(a_1)+(gf+fg)(a_2)+f^2(a_3)-(f+g)(0)} \\ \multicolumn{7}{c}{g^3(a_0)+(g^2f+gfg+fg^2)(a_1)+(gf^2+fgf+f^2g)(a_2)+f^3(a_3)-(f+g)^2(0)-4(f+g)(0)} \end{array}$$

where the factor of 4 appears because $f(a+b+c+\cdots) = f(a)+f(b)+f(c)+\cdots - kf(0)$ where $k$ is the number of $+$'s.

So in addition to powers of $f$ and $g$, we have a constant equal to

$$-\sum_{i=1}^{t-1}(2^i+i-2)(f+g)^{t-i}(0)$$

Except for a finite number of terms this sum will just depend on $(2^i+i-2) \bmod p$ for some $p$, which is eventually periodic, and $(t-i) \bmod q$ for some $q$, since $(f+g)^{t-i}$ is eventually periodic. So each term can be computed in constant time — and the whole sum in time linear in $t$.

Now that we know how to separate out the effect of $f(0)$ and $g(0)$, let's assume for the remainder of this section that $f(0) = g(0) = 0$; i.e., that $f$ and $g$ are homomorphisms. We then have a Green's function as before, with $P_t = \sum_{x=0}^{t} G_{x,t}(a_x)$, as we had for Abelian groups and semigroups; except now, $G$ is a kind of non-commuting Pascal's Triangle

$$\begin{array}{ccccccc} & & & 1 & & & \\ & & f & & g & & \\ & f^2 & & fg+gf & & g^2 & \\ f^3 & & f^2g+fgf+gf^2 & & fg^2+gfg+g^2f & & g^3 \end{array} \tag{5}$$

where each term represents a path to that site from the initial site.

This is a generalization of so-called *additive* CAs: rather than scalar coefficients, such as $a = 2b+3c \bmod 5$, we now have matrix coefficients (for instance, $2 \times 2$ matrices mod 2 if the underlying group is $Z_2^2$) which don't necessarily commute.

Once the $G_{x,t}$ are calculated, we can calculate the sum $P_t = \sum_x G_{x,t}a_x$ in serial time $t$ or $\log t$ in parallel. However, this isn't going to help us much unless the $G_{x,t}$ are fairly easy to calculate. If each one takes $t$ steps, for instance, then calculating all of them will take $t^2$ time, no better than complete simulation.

In (5), the number of terms in $G_{x,t}$ is $\binom{t}{x}$ and the total number of terms grows as $2^t$ — hardly what we need for an efficient algorithm. However, if $f$ and $g$ fulfill certain conditions, we'll be able to simplify (5) considerably. (Some of the following results will generalize to non-quasigroup isotopies, for instance if $a \bullet b = f(a) + g(b)$ where $+$ is a semigroup, or where $f$ and $g$ are non-invertible matrices and so are not one-to-one.)

## 5.1 $f$ and $g$ commute

If $fg = gf$, (5) becomes

$$\begin{array}{ccccccc} & & & 1 & & & \\ & & f & & g & & \\ & f^2 & & 2fg & & g^2 & \\ f^3 & & 3f^2g & & 3fg^2 & & g^3 \end{array}$$

or in general,

$$G_{x,t} = \binom{t}{x} f^{t-x} g^x.$$

Since $f$ and $g$ are one-to-one, $f^p = 1$ for some $p$, so $f^i$ only depends on $i \bmod p$. Similarly for $g$; so $f^{t-x}g^x$ can be calculated in constant time, and the sum $\sum G_{x,t}(a_x)$ can again be calculated in time $\mathcal{O}(t)$. In fact, this also works for CAs isotopic to an Abelian semigroup, where $f$ and $g$ are commuting homomorphisms.

This includes the usual "additive" CAs, where the underlying group is cyclic and $f$ and $g$ are scalars.

## 5.2 $f^2 = g^2 = 1$

If $f^2 = g^2 = 1$, for even time-steps (5) becomes

$$\begin{array}{ccccccc} & & & 1 & & & \\ & & 1 & h & 1 & & \\ & 1 & 2h & 2+h^2 & 2h & 1 & \\ 1 & 3h & 3+3h^2 & 6h+h^3 & 3+3h^2 & 3h & 1 \end{array}$$

where $h = fg + gf$. So even if $f$ and $g$ don't commute, all three functions that appear after two time-steps, $f^2$, $g^2$, and $fg + gf$, commute with each other; so we can use a $r = 1$ Pascal's Triangle to get

$$G_{x,t} = \sum_{k=0,2,4\ldots}^{t/2} \binom{t/2}{k} \binom{t/2-k}{(x-k)/2} h^k = \sum_{k=0,2,4\ldots}^{t/2} \frac{\frac{t}{2}!}{k!\,\frac{x-k}{2}!\,\frac{t-x-k}{2}!} h^k \qquad (6)$$

for even $x$. (For odd $x$, we sum over $k = 1, 3, 5 \ldots$ instead.) The first binomial coefficient chooses which $k$ steps in the path to that term in the triangle go straight down, gaining $k$ factors of $h$, and the second chooses which of the remaining steps go to the right. Their product is a *trinomial coefficient* $\binom{\frac{t}{2}}{k, \frac{x-k}{2}}$.

For odd $t$, we can easily derive $G_{x,t}$ from its two predecessors with even $t$. But how can we calculate the sum (6) efficiently?

We can use the same polynomial scaling as in section 3. If we define $G(h, x, t) = (1 + hx + x^2)^t$, we get

$$G(h, x, p) = (1 + hx + x^2)^p \equiv_p 1 + h^p x^p + x^{2p} = G(h^p, x^p, 1)$$

and

$$G(h, x, q) \equiv_p G(h^p, x^p, p^{m-1})$$

for a prime power $q = p^m$. So by decomposing the group into primes and prime powers as we did before, we can predict the CA in linear time (except since $G(h, x, 1)$ has three terms, $m$ is effectively 1 for primes and $m + 1$ for prime powers).

In fact, we can generalize this to the following:

**Theorem 4.** *Consider a CA separably isotopic to a Abelian group. If there exists a* $t$ *such that all the* $G_{x,t}$ *commute, then the CA can be predicted in serial time* $\mathcal{O}(t)$ *or parallel time* $\mathcal{O}(\log t)$*, so its prediction is in* $\mathbf{NC}_1$*.*

### 5.3 $f^2$ and $g^2$ commute with $f$ and $g$

If $gf^2 = f^2 g$ and $g^2 f = fg^2$, then we can group $f^2$s to the left and $g^2$s to the right. Every term in (5) then has the form $f^a (gf)^b g^c$, as in

$$\begin{array}{ccccccc} & & & 1 & & & \\ & & f & & g & & \\ & f^2 & & fg + gf & & g^2 & \\ f^3 & & 2f^2 g + fgf & & gfg + 2fg^2 & & g^3 \end{array}$$

Equation (6) still applies with a small modification: $f^2$ and $g^2$ don't disappear, but they can be moved aside, giving

$$G_{x,t} = \sum_{k=0,2,4\ldots}^{t/2} \frac{\frac{t}{2}!}{k! \, \frac{x-k}{2}! \, \frac{t-x-k}{2}!} f^{t-x-k} h^k g^{x-k}$$

for even $x$ and even $t$, where $h = fg + gf$ as before. Then we use the same algorithm as above, and again take serial time $\mathcal{O}(t)$ or $\mathcal{O}(\log t)$ in parallel.

## 5.4 $[f, g]$ commutes with $f$ and $g$

Suppose $f$ and $g$ don't commute with each other, but their *commutator* $[g, f] = gfg^{-1}f^{-1}$ does. Call this $h$. Then $gf = hfg$ and we can move $f$s to the left and $g$s to the right: for instance, $gfgf = h^3f^2g^2$.

Then (5) becomes

$$\begin{array}{ccccccc} & & & 1 & & & \\ & & f & & g & & \\ & f^2 & & (1+h)fg & & g^2 & \\ f^3 & & (1+h+h^2)f^2g & & (1+h+h^2)fg^2 & & g^3 \end{array}$$

We get increasing powers of $h$ from the left-hand predecessors, since $gf^{t-x}g^x = h^{t-x}f^{t-x}g^{x+1}$. Since each path from the initial site chooses $x$ places to turn right, and each one adds $t-x$ factors of $h$, we have

$$G_{x,t} = f^{t-x}g^x \sum_{k=0}^{x(t-x)} \Pi(k, t-x, x)h^k$$

where $\Pi(k, m, n)$ is the number of ways to partition $k$ indistinguishable objects into $n$ piles of 0 to $m$ objects each. For instance, $\Pi(5, 4, 3) = 4$: $(4, 1, 0)$, $(3, 2, 0)$, $(3, 1, 1)$, and $(2, 2, 1)$. $\Pi$ satisfies a number of identities, including $\Pi(k, m, n) = \Pi(k, n, m)$ and $\Pi(k, m, n) = \Pi(k-m, m, n-1) + \Pi(k, m-1, n)$.

Since $f$ and $g$ are automorphisms, $h$ is also, so $h^r = 1$ for some $r$. So in fact we just need to calculate

$$\sum_{k=0}^{r-1} \left( \sum_{j=0}^{\lfloor x(t-x)/r \rfloor} \Pi(k + jr, t-x, x) \right) h^k$$

mod a finite number of primes or prime powers as before. Perhaps the reader can find an efficient way to calculate this series.

## 5.5 Two examples: vector-valued CAs

We now show two *vector-valued* CAs — where the state is a vector and the coefficients $f$ and $g$ are matrices. In particular, our states will be two-component vectors mod 2 (the group $Z_2^2$), and our coefficients will $2 \times 2$ matrices mod 2 (the homomorphisms on that group).

The automorphisms (invertible matrices) of $Z_2^2$, are

$$\begin{pmatrix} 1 & 0 \\ 0 & 1 \end{pmatrix}, \begin{pmatrix} 1 & 1 \\ 1 & 0 \end{pmatrix}, \begin{pmatrix} 0 & 1 \\ 1 & 1 \end{pmatrix}, \begin{pmatrix} 0 & 1 \\ 1 & 0 \end{pmatrix}, \begin{pmatrix} 1 & 1 \\ 0 & 1 \end{pmatrix}, \begin{pmatrix} 1 & 0 \\ 1 & 1 \end{pmatrix}$$

which the reader can check are the identity, two period-3 rotations and three period-2 transpositions of the non-zero elements.

Since we already know what to do if $f$ and $g$ commute, we want a non-commuting pair. There are only two non-isomorphic ways to do this: a pair of distinct transpositions, and a transposition and a rotation. In the first case, for instance $f = \begin{pmatrix} 0 & 1 \\ 1 & 0 \end{pmatrix}$ and $g = \begin{pmatrix} 1 & 1 \\ 0 & 1 \end{pmatrix}$, we have $f^2 = g^2 = 1$ and we can use section 5.2 above.

### 5.5.1 Logarithmic scaling

The other case is more interesting. Let $f = \begin{pmatrix} 0 & 1 \\ 1 & 0 \end{pmatrix}$ and $g = \begin{pmatrix} 1 & 1 \\ 1 & 0 \end{pmatrix}$. Then $f^2 = g^3 = 1$, $fg + gf = f$, $g + 1 = g^2$, and $gf = fg^2$: so the Green's function is

$$\begin{array}{c}
1 \\
f \quad g \\
1 \quad f \quad g^2 \\
f \quad g^2 \quad 0 \quad 1 \\
1 \quad 0 \quad 1 \quad f \quad g \\
f \quad g \quad f \quad g^2 \quad f \quad g^2 \\
1 \quad f \quad g \quad 0 \quad 0 \quad 0 \quad 1 \\
f \quad g^2 \quad f \quad g^2 \quad 0 \quad 0 \quad f \quad g \\
1 \quad 0 \quad 0 \quad 0 \quad 1 \quad 0 \quad 1 \quad f \quad g^2
\end{array}$$

A space-time diagram is shown in figure 4. This Green's function has gaps similar to an Abelian group CA, but the gaps are skewed. For $t = 2^k$, the Green's function has $k + 2$ non-zero terms, located $0, 1, 2 \ldots 2^k$ sites from the right end of the row.

We can show this by induction: assume that (with $x$ counting from the right) the only non-zero $G_{x,t}$ are

$$G_{x,t} = \begin{cases} g & \text{if } x = 0 \\ f & \text{if } x = 1 \\ 1 & \text{if } x = 2^k,\ k > 0 \end{cases}$$

as we do for $t = 1$. Then the only non-zero $G_{x,2t}$ are

$$G_{x,2t} = \begin{cases} G_{0,t}^2 = g^2 & \text{if } x = 0 \\ G_{0,t}G_{1,t} + G_{1,t}G_{0,t} = fg + gf = f & \text{if } x = 1 \\ G_{0,t}G_{x,t} + G_{x,t}G_{0,t} + G_{x/2,t}^2 = 2g + 1 = 1 & \text{if } x = 2^k,\ k > 0 \end{cases}$$

and similarly if $G_{0,t} = g^2$.

We'll call this *logarithmic scaling* — it's more complex than an Abelian group CA, whose Green's function has only a constant number of non-zero terms when $t = p^k$. It has a corresponding scaling relation

$$P_t(a_0 \ldots a_t) = a_0 + a_t + P_{t/2}(a_{t/2} \ldots a_t)$$

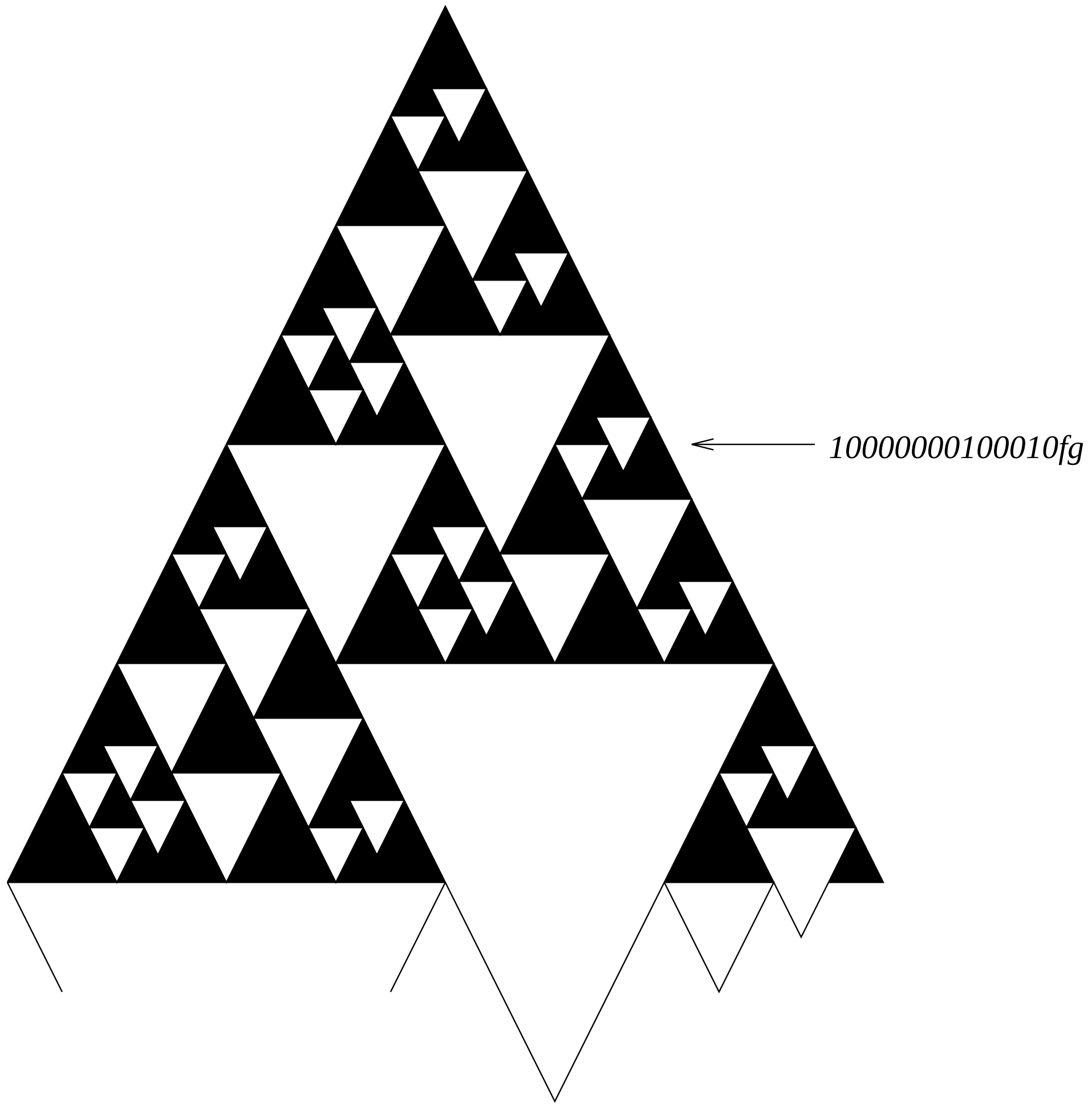

Figure 4: The Green's function for the logarithmic scaling CA. Non-zero elements are black, zero is white. When $t = 2^k$, the gaps leave only $k+2$ non-zero sites. The reader may recognize this as the Green's function for elementary CA rule 150, rotated 120° in space-time. More precisely, if we rotate the triangle of predecessors and successor $c = fa + gb$, we get a new rule $a = f'b + g'c$ where $f' = f^{-1}g$ and $g' = f^{-1}$; then $f'^2 = g'^2 = f'g' + g'f' = 1$, so after two time-steps we get rule 150, $P(a_0, a_1, a_2) = a_0 + a_1 + a_2$ on both components.

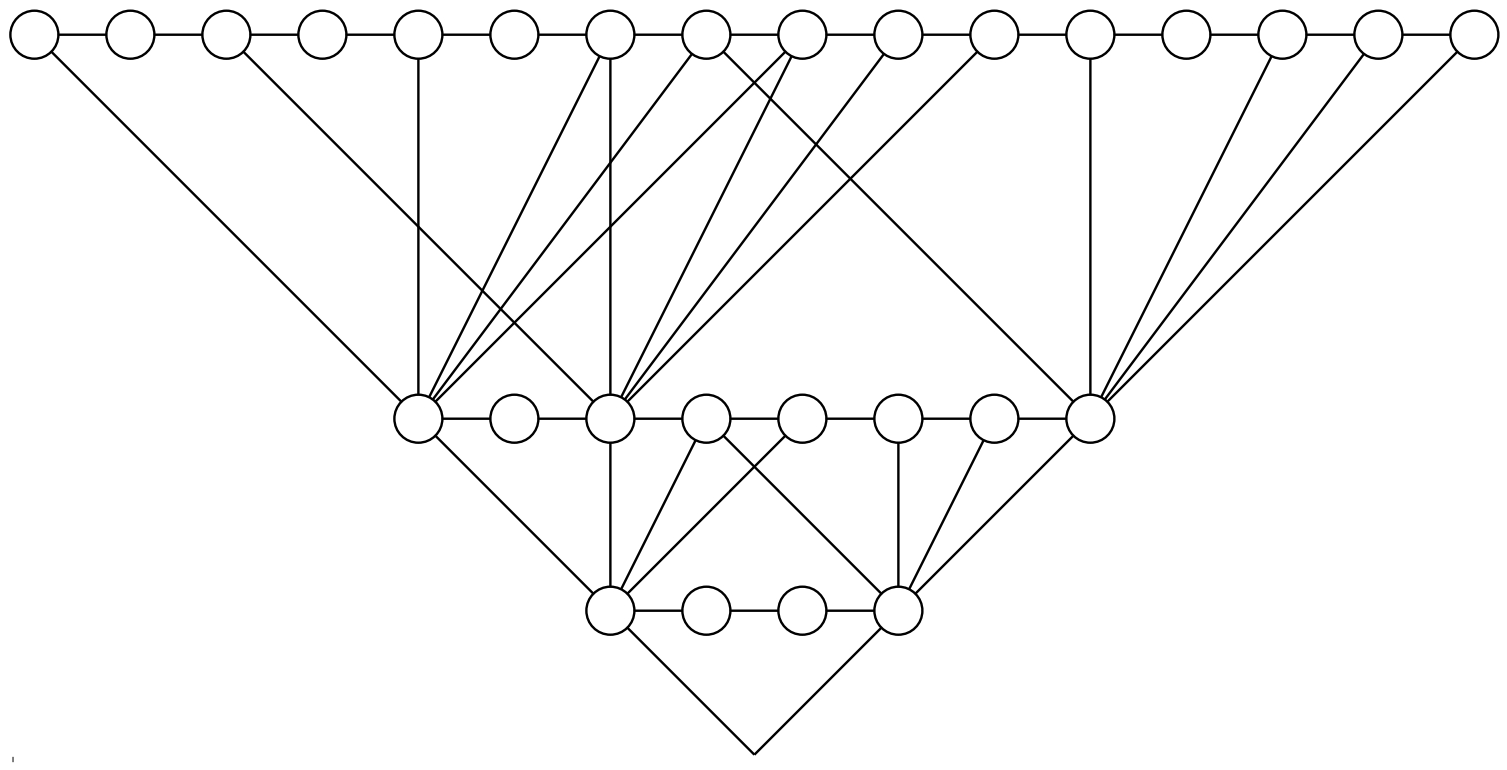

Figure 5: Scaling from above, coming down by decreasing powers of 2 from the initial row.

rather than the simple scaling $P_t(a_0, \ldots, a_t) = a_0 + a_t$ we encountered before.

So if $t$ is a power of 2, we can add up the contributions from the Green's function in $\mathcal{O}(\log t)$ serial time or $\mathcal{O}(\log \log t)$ in parallel (since it takes $\log n$ depth to add $n$ things). But what about for general $t$?

Using "scaling from above" as shown in figure 5, we start by jumping down $2^k$ steps from the initial row, where $k = \lfloor \log t \rfloor$ is the largest power of 2 in $t$. Each site in this row is the sum of $k+2$ initial sites; then coming down the next largest power of 2 from there, each site at $t = 2^k + 2^{k-1}$ depends on $k+1$ sites in the row at $t = 2^k$, and so on.

In the worst case where $t = 2^{k+1} - 1$, we have $k+1$ such rows; multiplying the number of sites in each row by the number of terms each site depends on in the row above it, we get a total time of

$$(k+2)2^k + (k+1)2^{k-1} + \cdots = \frac{d}{dr}\Big|_{r=2} \sum_{i=0}^{k+1} r^i = \mathcal{O}(k2^{k+1}) + \mathcal{O}(k^2) = \mathcal{O}(t \log t)$$

so we need serial time $\mathcal{O}(t \log t)$. In parallel, calculating this series of rows will take total depth

$$\log(k+2) + \log(k+1) + \cdots = \sum_{i=1}^{k} \log i = \mathcal{O}(k \log k) = \mathcal{O}(\log t \, \log\log t).$$

Alternately, we can use "scaling from below", as we did for scaling CAs before — jumping up the largest possible power of 2 from the bottom of the light-cone, then the next largest power of 2 above that, and so on. In the row $2^i$ above the bottom, we just need to know the sites whose positions are 0 or a power of 2 from the right, since that's where $G$'s non-zero terms are. In the row $2^{i-1}$ above that, we need to know the sites whose positions are a sum of two or fewer powers of 2, and so on. So, as shown in figure 6, the total number of sites we have to calculate in the worst case where $t = 2^k - 1$ is

$$1 + (k+1) + (\binom{k}{2} + k + 1) + \cdots + (2^k - 1) = \sum_{i=0}^{k-1} \sum_{j=0}^{i} \binom{k}{j} = \sum_{j=0}^{k} (k-j) \binom{k}{j}$$

$$= \frac{d}{dr}\Big|_{r=1} (1+r)^k = k2^{k-1} = (1/2) t \log t$$

so we get the same serial computation time up to a constant. In parallel, we'll do the same series of additions as with scaling from above, just in reverse order — giving the same parallel computation time.

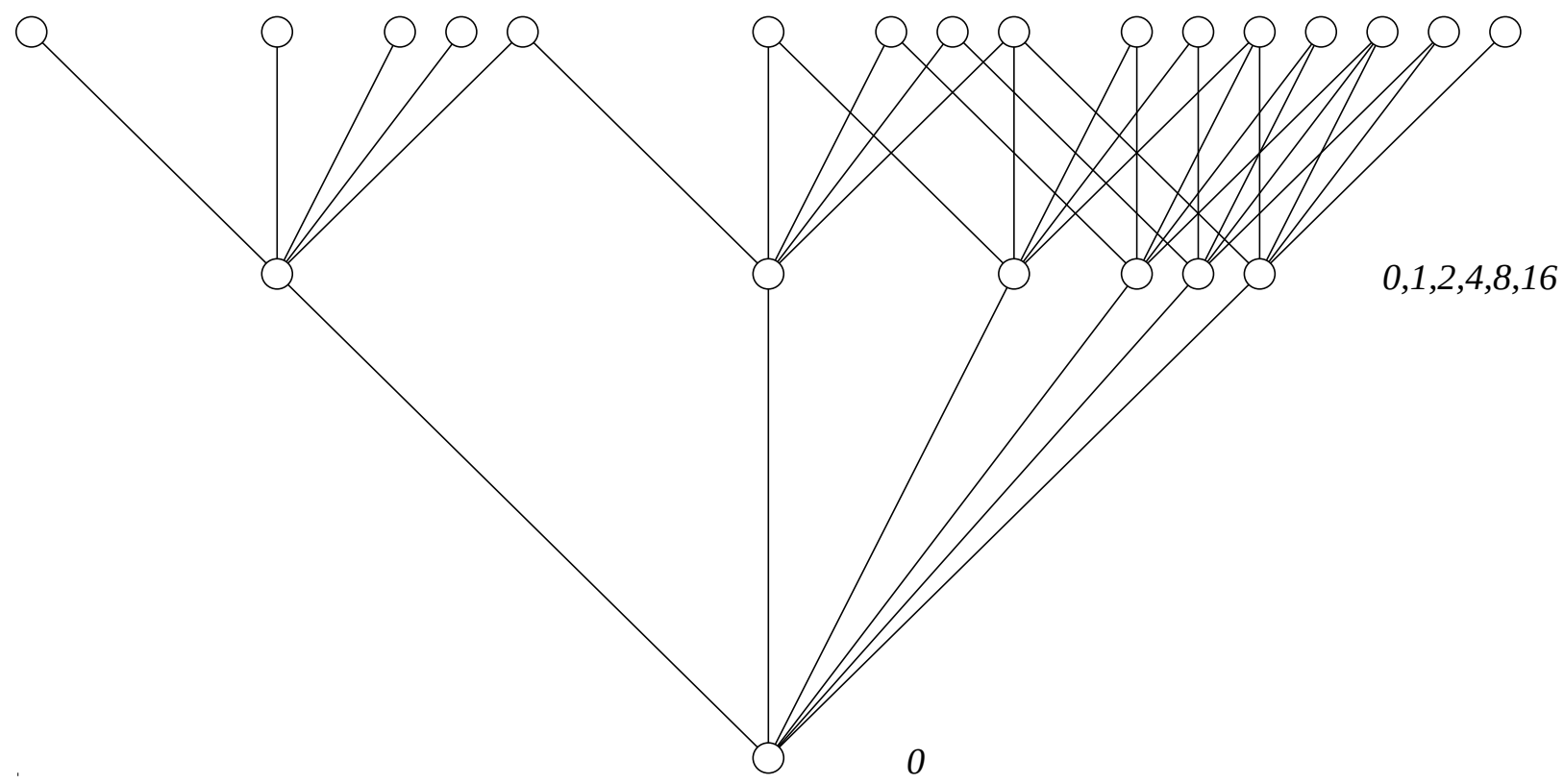

Figure 6: Scaling from below, jumping up by decreasing powers of 2 from the final site. The numbers show the positions (from the right) of the sites that need to be calculated.

The reader can confirm that the Green's function has this logarithmic scaling behavior whenever

$$\forall k : \, y_k^2 = 1 \text{ where } y_0 = f \text{ and } y_{k+1} = y_k g^{2^k} + g^{2^k} y_k \tag{7}$$

In this case, $y_k = f$ for all $k$ since $fg + gf = fg^2 + g^2 f = f$, and $f^2 = 1$. The non-quasigroup CA where $f = \begin{pmatrix} 1 & 1 \\ 0 & 1 \end{pmatrix}$ and $g = \begin{pmatrix} 0 & 0 \\ 1 & 1 \end{pmatrix}$ is another example, since $g^2 = g$, $fg + gf = f$, and $f^2 = 1$.

So we have

**Theorem 5.** *An additive CA whose coefficients satisfy equation (7) can be predicted in time $\mathcal{O}(t \log t)$ in serial or $\mathcal{O}(\log t \log\log t)$ in parallel (between $\mathbf{NC}_1$ and $\mathbf{NC}_2$).*

### 5.5.2 Morse sequence Green's functions

As another interesting example of a vector-valued CA on $Z_2^2$, let $f = \begin{pmatrix} 1 & 1 \\ 1 & 0 \end{pmatrix}$ and $g = \begin{pmatrix} 0 & 0 \\ 1 & 1 \end{pmatrix}$. Since $g$ isn't invertible, this isn't a quasigroup. We get the following Green's function:

$$\begin{array}{ccccccccccccccccc}
&&&&&&&&1&&&&&&&& \\
&&&&&&&f&&g&&&&&&& \\
&&&&&&f^2&&h&&g&&&&&& \\
&&&&&1&&1&&f&&g&&&&& \\
&&&&f&&h&&j&&h&&g&&&& \\
&&&f^2&&k&&f^2&&k&&f&&g&&& \\
&&1&&0&&1&&0&&f^2&&h&&g&& \\
&f&&g&&f&&g&&1&&1&&f&&g& \\
f^2&&h&&j&&h&&h&&h&&j&&h&&g
\end{array}$$

where $f^3 = 1$, $g^2 = g$, $fg+gf = h$, $g+f^2 = j$, and $g+1 = fh+gf = fh+gj = k$ where $h = \begin{pmatrix} 1 & 1 \\ 0 & 1 \end{pmatrix}$, $j = \begin{pmatrix} 0 & 1 \\ 0 & 0 \end{pmatrix}$, and $k = \begin{pmatrix} 1 & 0 \\ 1 & 0 \end{pmatrix}$.

Whenever $t = 2^k$, we see that $G_{x,t}$ consists of $f$ or $f^2$ (for odd or even $k$) followed by a *Morse sequence:*

$$hjhh\, hjhj\, hjhh\, hjhh\;\; hjhh\, hjhj\, hjhh\, hjhj \ldots$$

except that the last symbol is replaced by a $g$. This sequence is generated by the dilation operations

$$h \to hj,\; j \to hh,\; g \to hg$$

or alternately, by the requirement that the second half is identical to the first half except for the last symbol.

As with the previous example, we can show this by induction. Since $g = h + f = j + f^2$, when $t = 2^k$ we can write $G_t = f(1 + x^t) + M_t$ ($k$ even) or $G_t = f^2(1 + x^t) + M_t$ ($k$ odd), where $M_t$'s coefficients are all $h$s or $j$s; for instance, $M_4 = hx + jx^2 + hx^3 + hx^4$. Then

$$G_{2t} = G_t^2 = f^2(1 + x^{2t}) + (fM_t + M_t f)(1 + x^t) + M_t^2$$

or

$$M_{2t} = (fM_t + M_t f)(1 + x^t) + M_t^2$$

Let $m_{t,i}$ be the coefficient of $x_i$ in $M_t$, where $1 \le i \le t$. Since $fh + hf = fj + jf = h$, we have

$$m_{2t,2i-1} = h \text{ and } m_{2t,2i} = h + m_{t,i}^2$$

which, since $h^2 = 1$, $j^2 = 0$, and $h + 1 = j$, gives us the dilation rules $h \to hj$, $j \to hh$ above.

Again, we have a more subtle kind of scaling than before: this time,

$$P_t(a_0 \ldots a_t) = P_{t/2}(a_0 \ldots a_{t/2}) + P_{t/2}(a_{t/2} \ldots a_t) + a_t$$

This allows us to predict the CA in time $\mathcal{O}(t)$ in serial or $\mathcal{O}(\log t)$ in parallel if $t$ is a power of 2. What can we do for general $t$?

We can use the fact that these Morse sequences are almost periodic. First of all, $j = h + 1$, so if we subtract $h$ from every site we get

$$0100\,0101\,0100\,0100\,0100\,0101\,0100\,0101\ldots$$

The population of 0s is 2/3. Using the dilations $0 \to 01$, $1 \to 00$, we find that 0s map to blocks whose second half differs from the first half by one bit, while 1s map to blocks whose first and second halves are identical. Therefore, the sequence is period-1 except for 2/3 of the sites; period-2 except for 1/3 of the sites; period-4 except for 1/6 of the sites; and so on.

Now if we use scaling from above, we need to calculate rows of $l$ sites $2^k$ steps below the initial row where $l \leq 2^k$. We can calculate the first site in full, the second from the first by adding 2/3 of the initial sites, the third from the first (and the fourth from the second) by adding 1/3 of the initial sites, and so on: always shifting the largest possible power of 2 from a previously calculated site. This takes (serial) time

$$2^k(1 + (2/3) + (1/3) + (1/3) + (1/6) + (1/6) + (1/6) + (1/6) + (1/12) + \ldots)$$
$$= 2^k(1 + (2/3)\log l) \leq (2/3)k2^k + 2^k$$

as shown in figure 7.

Doing this for a series of rows descending by powers of 2 from the initial row, we get a worst-case leading behavior of $(4/3)t \log t$. (In parallel, we can calculate all $l$ sites in each row in time $\mathcal{O}(k)$, and then doing this for $\log t$ rows gives time $\mathcal{O}(\log^2 t)$.)

However, this doesn't quite work. When we calculate a new site in a row from a site $2^m$ to the left by shifting the Green's function, we also have to add contributions from $2^m$ new initial sites on the right (and subtract $2^m$ of them from the left). Of course, we can use the same shifting trick to calculate these $2^m$ sites from previously calculated sums of $2^m$ sites, and so on; but even when we use these techniques to their best advantage, we still find the time it takes to calculate the $2^k$th site in a row, even if the previous $2^k - 1$ sites are known, is proportional to

$$\sqrt{2k}\, 2^{\sqrt{2k}}$$

Each row takes at least $2^k$ times this to calculate, and then summing over successive values of $2^k$ we get a leading behavior of

$$t\sqrt{2\log t}\, 2^{\sqrt{2\log t}}$$

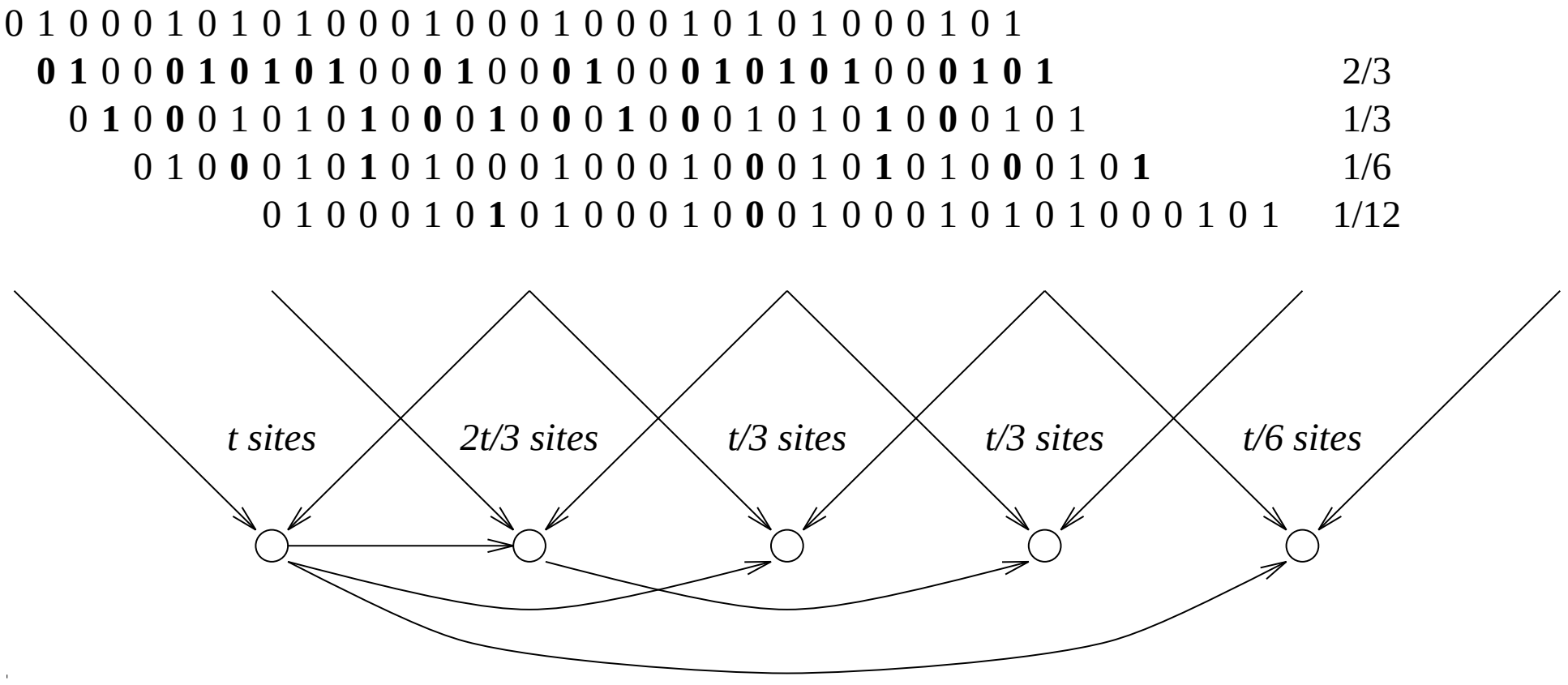

Figure 7: The Morse sequence is almost periodic. Shifting by a power of 2 changes an exponentially decreasing number of bits (shown here in bold). Therefore, sites can be calculated from others in the same row by adding contributions from an exponentially decreasing number of initial sites.

This is an interesting function, bigger than $t \log^k t$ for any $k$, but less than $t^{1+\epsilon}$ for any $\epsilon > 0$.

However, as fun as this algorithm is to work out, a more general method is in fact more efficient, as we will show next.

## 5.6 A general method

Is there a method that works for all quasigroups separably isotopic to an Abelian group? The polynomial method we used before works perfectly well with matrix coefficients — just let $G_{i,t}$ be the coefficient of $x^i$ in $G_1^t$ where $G_1 = (f + gx)$. In fact, this works just as well if $f$ and $g$ are homomorphisms, so the CA doesn't need to be a quasigroup.

Adding the sum $\sum_x G_{x,t} a_x$ only takes linear serial time or logarithmic parallel time. So how fast can we calculate $(f + gx)^t$?

In order to raise a polynomial $G_1$ to the $t$th power, we can iterate between two operations: squaring it, which doubles $t$, and multiplying by $G_1$, which increments $t$ by 1. The latter can be done in linear time; the former is more difficult, and we'll need to do it up to $\log t$ times.

Multiplying $n$th-order polynomials in one variable is essentially the same as multiplying $n$-digit numbers, except that no carrying of digits takes place. The fastest known serial algorithms [17] take time $\mathcal{O}(n \log n \, \log \log n)$ on a multitape Turing machine, $\mathcal{O}(n \log n)$ on a RAM, and $\mathcal{O}(n)$ on a *storage modification machine*, a machine which can create new registers with pointers to each other. (Multiplying polynomials whose coefficients are $n \times n$ matrices, for fixed $n$, just adds a multiplicative constant to the computation time in serial and an additive constant in parallel.)

Iterating all of these $\log t$ times on successively doubling values of $n$ gives $n \propto t$ as the leading behavior, so:

**Theorem 6.** *Any CA which is additive on an Abelian group with matrix coefficients (including CAs which are separably isotopic to an Abelian group) can be predicted in serial time $\mathcal{O}(t \log t \log\log t)$ on a multi-tape Turing machine, $\mathcal{O}(t \log t)$ on a RAM, or $\mathcal{O}(t)$ on a storage modification machine.*

In parallel, $n$-digit numbers can be multiplied in time $\mathcal{O}(\log n)$ by $\mathcal{O}(n^2)$ processors, so summing over $n = 2^k$ for $k \leq \log t$ we get time $(1/2)\log^2 n$. So

**Theorem 7.** *Any CA which is additive on an Abelian group with matrix coefficients can be predicted in parallel time $\mathcal{O}(\log^2 t)$ by $\mathcal{O}(t^2)$ processors, so its prediction is in $\mathbf{NC}_2$.*

# 6 Non-Abelian groups

If a group is non-Abelian, we can no longer use the Green's function formalism: each $a_i$'s contribution is entangled with the others', so $P_t$ is no longer a linear function of the $a_i$.

Unforunately, the isotopy approach can't help us either. Since groups are isomorphic if they're isotopic [12], a non-Abelian group can't be isotopic to an Abelian one. However, for certain non-Abelian groups we can still find an efficient solution.

## 6.1 The Quaternions $Q_8$

The *Quaternion group* $Q_8$ is the group $\{1, -1, i, -i, j, -j, k, -k\}$ with the following relations:

$$i^2 = j^2 = k^2 = -1$$

$$ij = k, jk = i, ki = j$$

$$ji = -k, kj = -i, ik = -j$$

and 1 and $-1$ act as we would expect ($(-1)i = i(-1) = -i$, etc.)

Note that $Q_8$ is "almost Abelian" in that $ab = \pm ba$, or $[a, b] = \pm 1$ where $[a, b] = aba^{-1}b^{-1}$ is the *commutator* of $a$ and $b$. The commutator describes to what extent $a$ and $b$ don't commute: we can use it to re-arrange variables, as in $ab = [a, b]ba$.

The *center* of a group is the subgroup of all elements that commute with every element, in this case $\{1, -1\}$. All commutators are in the center; so are all squares, since $a^2 = \pm 1$ for all $a$. So $[a, b] = 1$ if $a$ or $b$ is a commutator or a square. We also have $a^4 = 1$ and $[a, b]^2 = 1$, and we can derive identities like $[a, b] = a^2 b^2 (ab)^2$ and $[ab, cd] = [a, c][a, d][b, c][b, d]$.

The upshot of all this is

$$\begin{array}{ccccccccc}
a_0 & & a_1 & & a_2 & & a_3 & & a_4 \\
& a_0a_1 & & a_1a_2 & & a_2a_3 & & a_3a_4 & \\
& & a_1^2a_0a_2 & & a_2^2a_1a_3 & & a_3^2a_2a_4 & & \\
& & & a_1^2a_2^2a_0a_2a_1a_3 & & a_2^2a_3^2a_1a_3a_2a_4 & & & \\
& & & & a_2^2[a_1,a_3]a_0a_4 & & & &
\end{array}$$

and in general, if $t = 2^k$ for some $k$,

$$P_t(a_0, \ldots, a_t) = a_{t/2}^2 \left( \prod_{i=1}^{t/2-1} [a_i, a_{t-i}] \right) a_0 a_t. \tag{8}$$

We can prove this by induction: if it's true for $t/2$, then

$$\begin{aligned}
P_t = P_{t/2}(P_{t,2}(a_0, \ldots, a_{t/2}), P_{t/2}(a_1, \ldots, a_{t/2+1}), \ldots P_{t/2}(a_{t/2}, \ldots, a_t)) \\
= (a_{t/4}a_{3t/4})^2 \left( \prod_{i=1}^{t/4-1} [a_i a_{t/2+i}, a_{t/2-i}a_{t-i}] \right) \\
\times a_{t/4}^2 \left( \prod_{i=1}^{t/4-1} [a_i, a_{t/2-i}] \right) a_{3t/4}^2 \left( \prod_{i=1}^{t/4-1} [a_{t/2+i}, a_{i-1}] \right) a_0 a_{t/2}^2 a_t
\end{aligned}$$

which simplifies to (8).

So except for the sign, this CA acts just like $Z_2^2$ with $(0,0) = \pm 1$, $(0,1) = \pm i$, $(1,0) = \pm j$, and $(1,1) = \pm k$ (since, algebraically, $Q_8/\{\pm 1\} = Z_2^2$) with the sign determined by the product of commutators in (7).

So if $t$ is a power of 2, $P_t$ can be calculated in linear time. For other $t$, we can go up by powers of 2 from below. The sign of $s$ is the product of the row of commutators times the signs of its predecessors $2^k$ steps above; their signs are their rows of commutators times the signs of their predecessors $2^{k-1}$ steps above them, and so on. So we need to calculate rows of intermediate products as shown in figure 8. Since commutators and squares can be removed from a commutator's arguments (since they commute with everything), we only need to know these rows up to a sign, i.e. the $Z_2^2$ part of them. So we need to make a short diversion.

### 6.1.1 Calculating a row of a scaling CA

As we discussed in section 3, $Z_p$ with $p$ prime scales with scaling ratio $p$, so for instance $Z_2$ scales with ratio 2. A single site at the bottom of a light-cone can be calculated in linear serial time as shown in figure 2. But now suppose we want to calculate a row $s_0, s_1, \ldots s_{l-1}$ of $l$ sites $t$ time-steps down from the initial conditions. In parallel, we can calculate all of them at once in time $\mathcal{O}(\log t)$; but to do it efficiently in serial requires more thought.

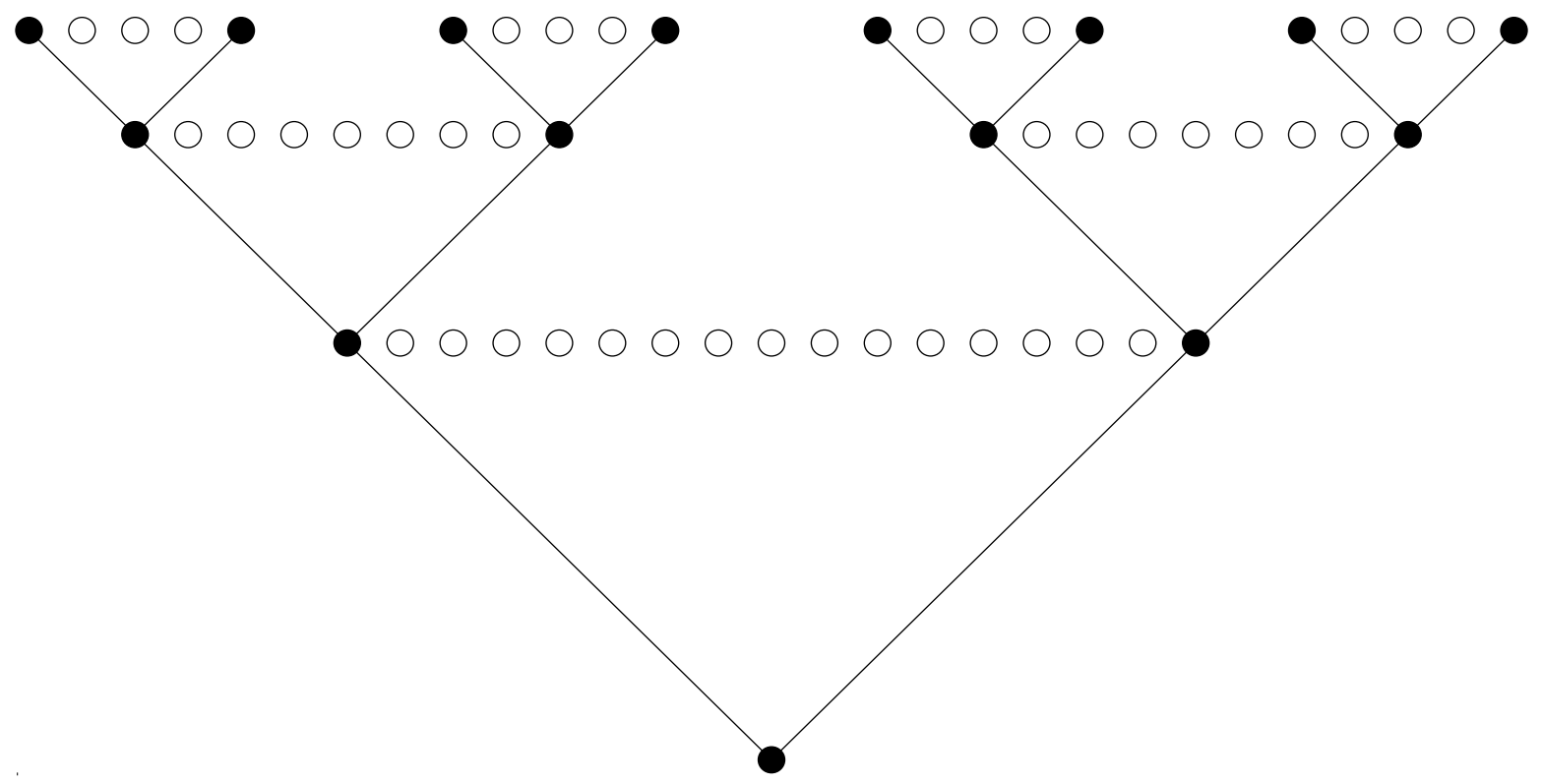

Figure 8: The intermediate states that need to be calculated to predict the Quaternion CA. The hollow circles just need to be calculated up to a sign, since only their squares or commutators affect the final state.

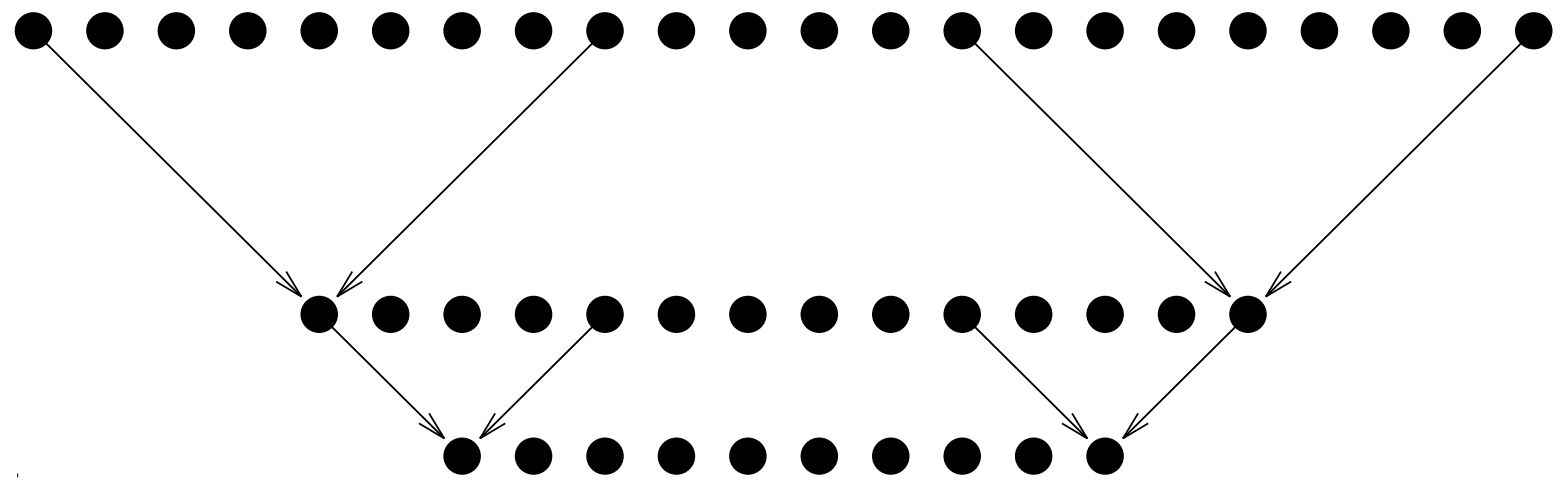

Figure 9: Using scaling from above to calculate a row of a scaling CA.

If we calculate each one independently using this method, the serial time is $lt$. But the area of the space-time trapezoid above the target row is $(l+t/2)t$, so unless $l \ll t$ this isn't any better than direct simulation. Luckily, the neighboring sites in the target row are not independent, so we can save time by re-using information.

There are two ways to do this: "scaling from above" is shown in figure 9. Calculate an entire row $2^k$ below the initial conditions, and then take the next power of 2 down and so on until we reach the target row. The computation time is then

$$l\#_1(t) + \sum_{i=1}^{\#_1(t)} (t - \sum_{j=1}^{i} 2^{k_j(t)}) < l\#_1(t) + t - \log t$$

where $\#_1(t)$ is the number of 1s in $t$'s digit sequence, and $k_j(t)$ is the position of the $j$th 1, starting from the most significant. So the leading behavior is $l\#_1(t) + t$, which is $l \log t + t$ in the worst case (namely, if $t = 2^k - 1$).

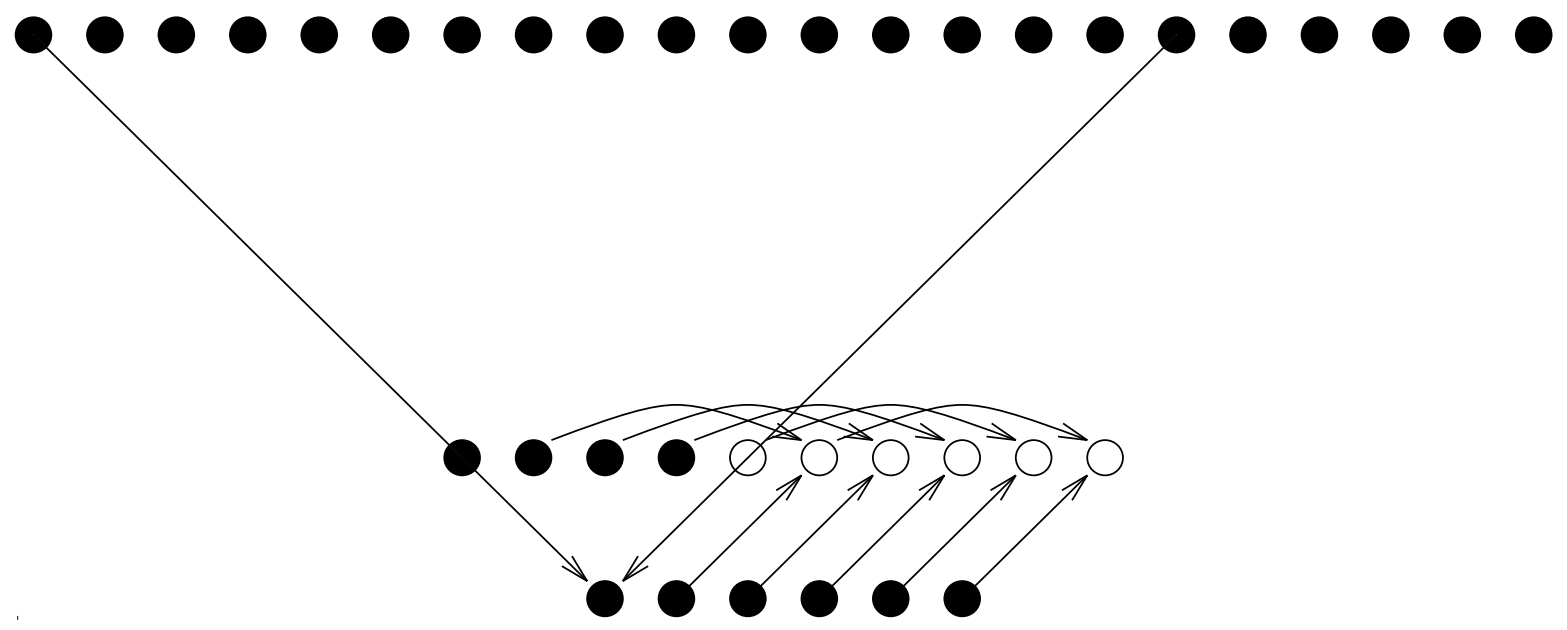

Figure 10: Another way of calculating a row of adjacent sites in a scaling CA: jump down to a $t + \Delta t$ with a sparse Green's function (i.e., a sum of a few powers of 2) and then reverse the CA rule to get the row at $t$.

This works well if $t$ has just a few 1s in its digits, and $G_{x,t}$ is sparse (zero for most $x$). But if $t$ has a lot of 1s, $G_{x,t}$ will have a lot of overlap with $G_{x+1,t}$ so $s_i$ and $s_{i+1}$ are closely related. For instance, if $t = 2^k - 1$, $G_{x,t} = 1$ for all $0 \le x \le t$ and $s_{i+1} = s_i + a_{i+t+1} - a_i$.

This is equivalent to calculating the row immediately below this one, at $t + 1 = 2^k$ whose Green's function only has two non-zero entries, and then reversing the CA to get $s_{i+1}$ from $s_i$ and their successor.

In general, suppose we find some $\Delta t$ such that $G$ is sparse for both $t + \Delta t$ and $\Delta t$. Then we calculate the row at $t + \Delta t$ and reverse $\Delta t$ steps of the CA to get the row at $t$. (We have to calculate the first $\Delta t$ of the row at $t$ before we can get going; then we use reversal to get the remaining $l - \Delta t$ of them.) This takes time

$$\Delta t \, \#_1(t) + (l - \Delta t)(\#_1(t + \Delta t) + \#_1(\Delta t)) + 2(t + \Delta t).$$

So for a fixed $\Delta t$ and increasing $l$, the leading behavior of this algorithm will be less than the first if $\#_1(t + \Delta t) + \#_1(\Delta t) < \#_1(t)$, i.e. if $t$ is the difference between two numbers with fewer total 1s than $t$ has. This is shown in figure 10.

This is true whenever $t$ has a long string of 1s in its digit sequence, e.g. $101110 = 110000 - 10$. So for $t = 101010\ldots$, or roughly $t = 2^k/3$, this algorithm can't improve matters. Since $\#_1(t) = (1/2) \log t$ then, both algorithms take serial time proportional to $l \log t + t$.

Note that we have not shown this is a lower bound — only that these two algorithms have this behavior in the worst case. Scaling ratios other than 2 simply change these results by a constant.

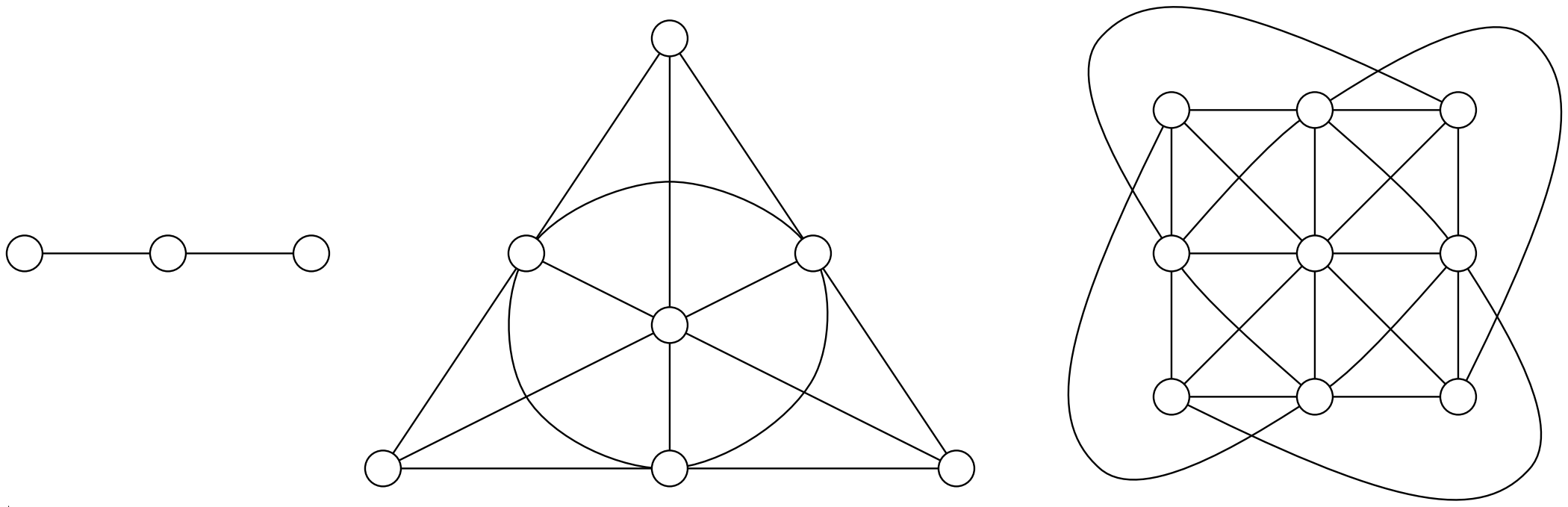

Figure 11: The Steiner triple systems for $n = 3$, 7, and 9.

### 6.1.2 Back to the Quaternions

Using this formula, we return to figure 7. The time to calculate all the rows (ignoring the term linear in $t$ because $l_i > t_i$) is at most

$$\sum_{i=1}^{\#_1(t)} 2^{i-1} l_i \, \#_1(t_i) = \sum_{i=1}^{\#_1(t)} 2^{i-1} 2^{k_i(t)} (\#_1(t) - i) \leq 2^{k_1} \#_1(t)^2/2 < t(\log t)^2$$

where the inequality is a proportionality in the worst case. So the Quaternion CA can be predicted in serial time $\mathcal{O}(t \log^2 t)$. Again, we can't rule out the possibility of a cleverer algorithm, but we conjecture that this is a lower bound.

In parallel, we can calculate all the rows of commutators in $\mathcal{O}(\log t)$ steps, multiply them together to get their products in another $\mathcal{O}(\log t)$ steps, and multiply all those products to get the sign of the final state in another $\mathcal{O}(\log t)$ steps — for a total which is still $\mathcal{O}(\log t)$. So in parallel, the Quaternion CA can be predicted just as quickly as one based on an Abelian group (although the number of processors grows as $\mathcal{O}(t^2)$ instead of $\mathcal{O}(t)$).

**Theorem 8.** *The Quaternion CA can be predicted in* $\mathcal{O}(t \log^2 t)$ *in serial or* $\mathcal{O}(\log t)$ *in parallel (with* $\mathcal{O}(t^2)$ *processors), so its prediction is in* $\mathbf{NC}_1$.

# 7 Steiner systems

In addition to groups, there are other algebraic structures on which a CA could be based. A *Steiner triple system* [14] is a set $S_{3,n}$ with a set of three-element subsets or *triples* such that for any two distinct elements $s_1$, $s_2$ there is a unique triple containing them.

A simple counting argument shows that the number of elements $n$ in a Steiner triple system must satisfy $n \bmod 6 = 1$ or 3, and the total number of triples is $n(n-1)/6$. There are in fact systems $S_{3,n}$ of every possible size. For $n = 3$, 7 and 9 the system is unique: $S_{3,3}$ is a single triple $\{1,2,3\}$, $S_{3,7}$ is a cyclically symmetric system

$$\{1,2,4\},\{2,3,5\},\{3,4,6\},\{4,5,7\},\{5,6,1\},\{6,7,2\},\{7,1,3\}$$

and $S_{3,9} = S_{3,3} \times S_{3,3}$, with triples of the form $\{(x_1,y_1),(x_2,y_2),(x_3,y_3)\}$ where the $x_i$ are either all different or all the same, and similarly for the $y_i$. These are shown in figure 11. There are two non-isomorphic $S_{3,13}$, one of which is cyclic (with 13 copies each of the triples $\{1,2,5\}$ and $\{1,3,8\}$) and 80 non-isomorphic $S_{3,15}$.

## 7.1 Squags

To turn a Steiner system into an algebra, we can define a binary operation $*$ on $S$: let $s_1 * s_2$ be the third element in the unique triple containing $s_1$ and $s_2$. Since the triples are unordered, this operation is totally symmetric: if $a * b = c$, then $c * b = a$ and similarly for all other permutations; and $(a * b) * b = a$.

The only remaining question is what value to assign to $a * a$. If we choose to make the elements *idempotent*, $a * a = a$, we get a kind of quasigroup called a *squag* [14].

It's easy to see that squags are not associative, since if $a * b = c$ then $a*(b*c) = a*a = a$ but $(a*b)*c = c*c = c$. Our only method so far of dealing with non-associative quasigroups is to make them isotopic to Abelian groups. Under what circumstances can a squag be isotopic to a group?

Suppose $a * b = f(a) \bullet g(b)$ where $\bullet$ is a group operation with identity $e$. Let $f'(a) = f(a) \bullet f(e)^{-1}$ and $g'(b) = f(e) \bullet g(b)$; then $a * b = f'(a) \bullet g'(b)$ and $f'(e) = e$. So we can assume $f(e) = e$, and $e = e * e = e \bullet g(e) = g(e)$, so $g(e) = e$ also.

Now $g(a) = e \bullet g(a) = e * a$, and $f(a) = f(a) \bullet e = a * e$. But since $*$ is commutative, $f(a) = g(a)$. Then $a = a * a = f(a) \bullet f(a) = f(a)^2$; similarly $f(a) = f(f(a))^2$, so $a = f(f(a))^4$. But $f(f(a)) = ((a * e) * e) = a$, so $a = a^4$. So $a^3 = 1$, and $f(a) = a^2 = a^{-1}$.

Now $a \bullet b = a^{-1} * b^{-1} = b^{-1} * a^{-1} = b \bullet a$, so $(S, \bullet)$ is in fact Abelian. Since $a^3 = 1$ for all $a$, $(S, \bullet)$ must be $Z_3^k$ for some $k$ and $n = 3^k$. Prediction is easy, since $a * b = (ab)^{-1}$: $P_t = P_{Z_3^k,t}$ for even $t$ and $P^{-1}_{Z_3^k,t}$ for odd $t$.

But this only works for Steiner systems that are some power of $S_{3,3}$. For instance, the $S_{3,7}$ squag is not isotopic to any group, and it's unclear how, or whether, it can be predicted efficiently.

## 7.2 Sloops

Instead of making the elements idempotent, we can add an identity $e$ and make them *nilpotent*, $a*a = e$. Then $(S,*)$ is a quasigroup with an identity, or a *loop*; this particular kind is called a *sloop* [14].

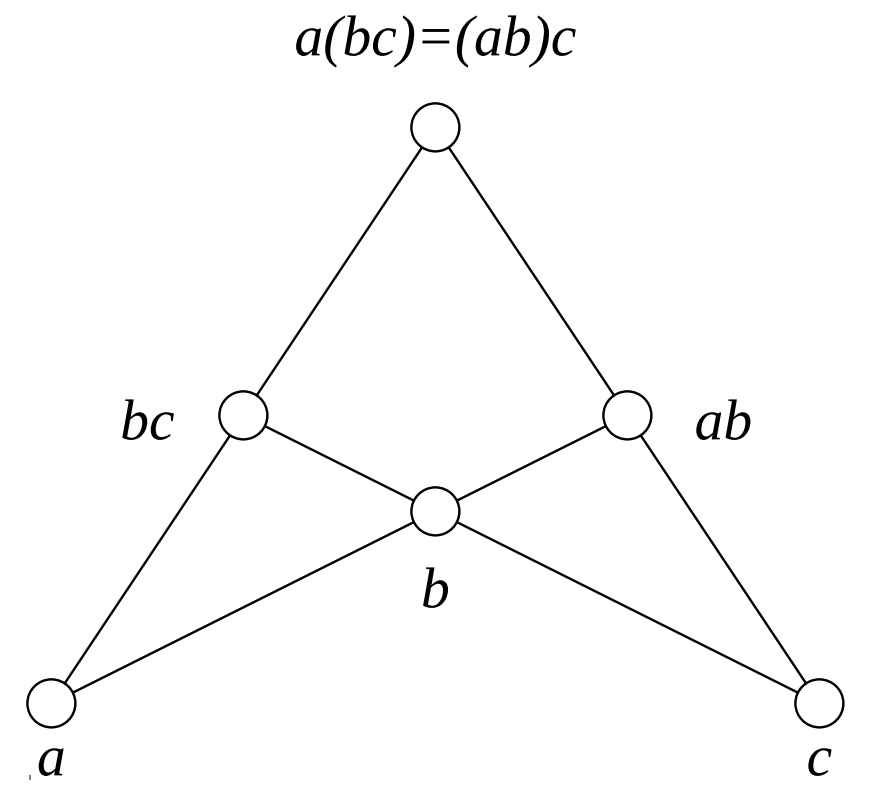

Figure 12: A graphical representation of associativity, on three distinct elements that do not form a triple (i.e., are not collinear).

Now if a loop is isotopic to a group, it's isomorphic to it [12], so the only sloops that can be isotopic to groups are those that are groups themselves. They're Abelian, and $a * a = e$, so they're $Z_2^k$ for some $k$ and the Steiner system has $n = 2^k - 1$. The elements of $S_{3,n}$ are non-zero vectors of $k$ components that sum to zero mod 2 (surprisingly, setting $k = 3$ gives the same $S_{3,7}$ as the cyclic structure shown above). These are also precisely those Steiner systems that are associative on any three distinct elements that do not form a triple, shown graphically in figure 12.

The smallest sloop that cannot be handled this way is the 10-element sloop formed by adding an identity to $S_{3,9}$. Again, it's not clear whether there's an efficient algorithm for it.

However, sloops and squags act the same on a restricted set of initial conditions: namely, those in which substrings of the form $aa$ never occur, and whose successors have the same property. This means $P_t(a_i, \ldots, a_{i+t}) \neq P_t(a_{i+1}, \ldots, a_{a+t+1})$ for all $t$, so

$$a_i \neq a_{i+1}, a_i \neq a_{i+2}, a_i \neq a_{i+1}(a_{i+2}a_{i+3}),$$

$$a_i \neq a_{i+1}((a_{i+1}a_{i+2})((a_{i+2}a_{i+3})(a_{i+3}a_{i+4}))), \ldots$$

We conjecture that, for $n > 3$, these constraints leave a non-empty set of initial conditions on which we can handle $S_{3,7}, S_{3,9}, S_{3,15}$, and others whose sloops or squags are isotopic to $Z_2^k$ or $Z_3^k$. It would be nice if deviations from this set of initial conditions could be handled as defects on a predictable background (such as in [8]).

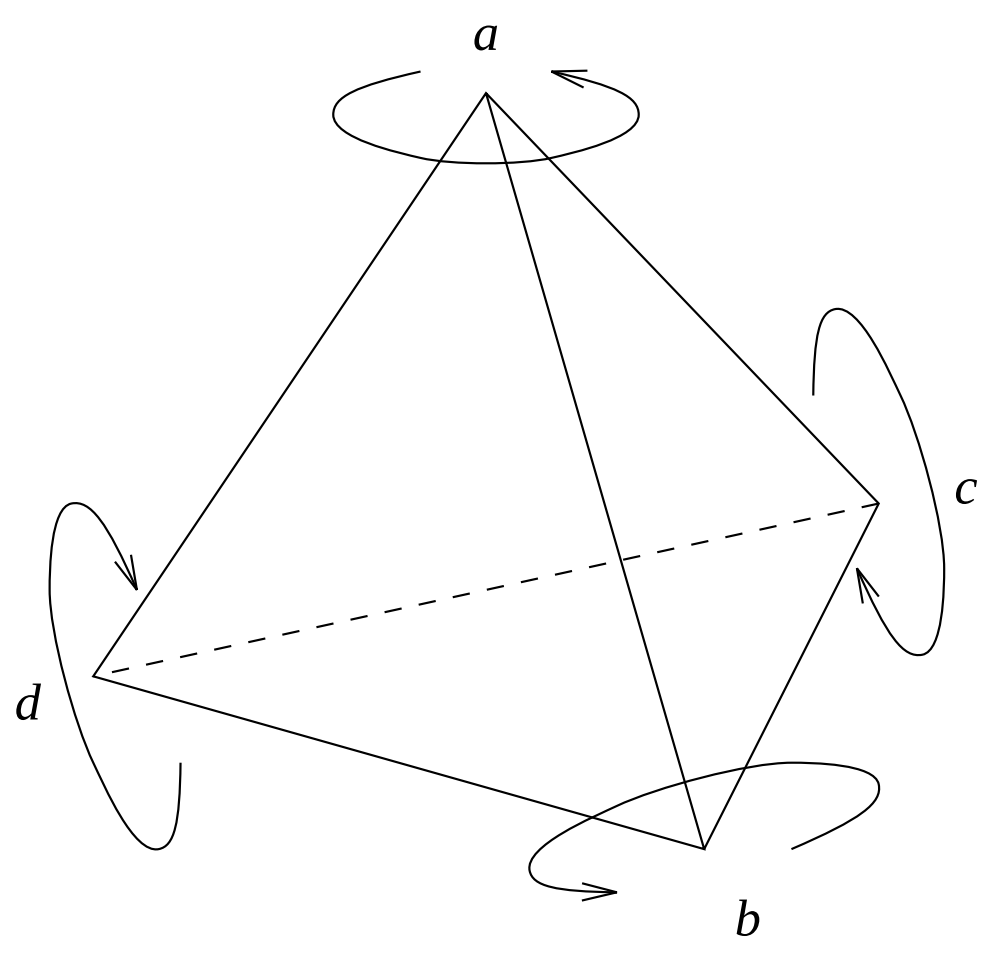

Figure 13: Multiplication in a Stein quasigroup. Each element leaves itself fixed (is idempotent) and rotates the other three.

### 7.3 Quadruple systems and Stein quasigroups

We can generalize the idea of a Steiner triple system in several ways. A *Steiner quadruple system* is a set $S_{4,n}$ with four-element subsets or *quadruples*, such that every distinct pair of elements is contained in a unique quadruple. The number of elements $n$ satisfies $n \bmod 12 = 1$ or 4, and the number of quadruples is $n(n-1)/12$. $S_{4,4}$ is a single quadruple; $S_{4,13}$ is cyclically symmetric with 13 copies of $(1, 2, 5, 7)$, and $S_{4,16}$ is $S_{4,4}^2$.

A Steiner quadruple system can be made into an quasigroup by assigning an orientation or parity to each quadruple. From a quadruple $(a, b, c, d)$ we define $a * b = c$ and $b * a = d$, and then use permutations of even parity to get the rest of the multiplication table:

| $*$ | $a$ | $b$ | $c$ | $d$ |
|---|---|---|---|---|
| $a$ | $a$ | $c$ | $d$ | $b$ |
| $b$ | $d$ | $b$ | $a$ | $c$ |
| $c$ | $b$ | $d$ | $c$ | $a$ |
| $d$ | $c$ | $a$ | $b$ | $d$ |

with each element idempotent and rotating the other three, just as if we rotated a tetrahedron around one of its vertices as shown in figure 13. This *Stein quasigroup* [14] satisfies the relations $(a * b) * b = b * a$ and $(b * a) * b = a$.

When is it isotopic to a group? Again, assume $a * b = f(a) \bullet g(b)$ where $\bullet$ has identity $e$ and $f(e) = e$. Then $e = e * e = e \bullet g(e) = g(e)$, so $g(e) = e$ also.

Now $f(a) = f(a) \bullet e = a * e$ and $g(a) = e \bullet g(a) = e * a$. But $f(g(a)) = (e * a) * e = a$, so $g = f^{-1}$ and $f$ and $g$ commute. Moreover, $f^3(a) = ((a * e) * e) * e = a$,

so $f^3 = 1$ and $g = f^2$.

But we also have

$$a \bullet a = (e * a) * (a * e) = (e * a) * ((e * a) * a) = a * (e * a) = e$$

so $a^2 = e$ for all $a$. So the group must be $Z_2^k$ for some $k$; $2^k \bmod 12 = 4$ for $k \geq 2$ even and 8 for $k \geq 3$ odd, so $k$ must be even and $n$ is a power of 4.

We also need to check that $f$ is a homomorphism on $\bullet$. If we replace $a$ with $f(a)$ and $b$ with $g(b)$, then $*$ must have an interesting *self-distributive* property (for left or right multiplication)

$$e * (a * b) = (e * a) * (e * b).$$

But this follows from another fact: since $\bullet$ is Abelian, $a \bullet b = b \bullet a$ and $(e * a) * (b * e) = (e * b) * (a * e)$. But then (omitting $*$)

$$a(ec) = (e(ae))((ce)e) = (e(ce))((ae)e) = c(ea).$$

Since $S$ is isotopic to a group, every element of $S$ is isomorphic to every other; specifically,$\bullet$ is isomorphic to another operation $\circ$ with identity $b$ instead of $e$, $a \circ c = a \bullet c \bullet b^{-1}$. So we can replace $e$ with any element $b$, and get *Abel-Grassmann's law*

$$a(bc) = c(ba).$$

This identity is shown graphically in figure 14 (it holds automatically if $a$, $b$ and $c$ are in the same quadruple). But then since $(ab)(cd) = d(c(ab)) = d(b(ac)) = (ac)(bd)$, we also have the *medial identity*

$$(ab)(cd) = (ac)(bd)$$

from which self-distributivity follows:

$$e(ab) = (ee)(ab) = (ea)(eb).$$

Self-distributivity means that multiplication by $e$, or any other element, is an automorphism on $S$: in particular, it sends each quadruple to another quadruple, either the same one or one disjoint from it. So we can choose a quadruple, and group all of $S$ into its images, giving us another Steiner system (quadruples of quadruples) with $n/4$ elements. It has the same properties, so we can continue the process inductively until we get to a single quadruple: so $S = (S_{4,4})^k$ for some $k$. (We can use a similar argument to show that the squag above is in fact a power of $S_{3,3}$.) But this leaves us with no algorithm for $S_{4,13}$.

In any case, we conclude this section with

**Theorem 9.** *If a Stein quasigroup is isotopic to a group, it is $S_{4,4}^k$ for some $k$ and $f = g^{-1}$; if a squag is isotopic to a group, it is $S_{3,3}^k$ for some $k$ and $f = g$. In either case, $f$ and $g$ commute and so it can be predicted in $\mathcal{O}(t)$ serial time or $\mathcal{O}(\log t)$ parallel time, and its prediction is in $\mathbf{NC}_1$.*

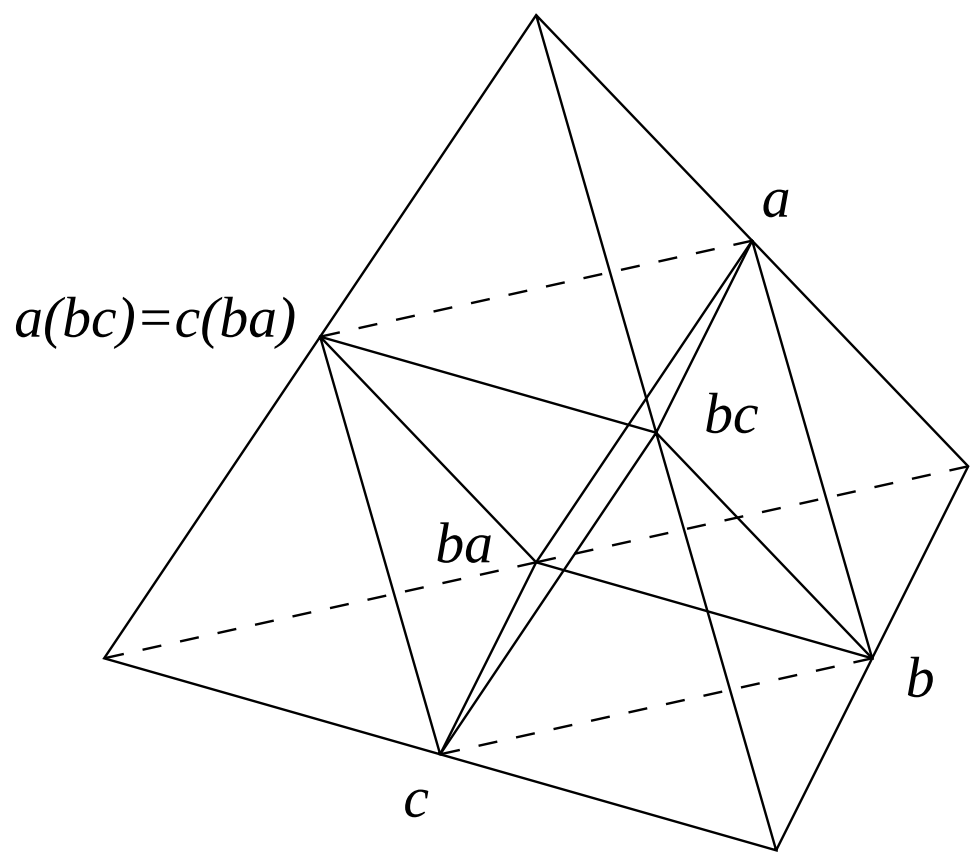

Figure 14: A graphical representation of Abel-Grassmann's law $a(bc) = c(ba)$ in a Stein quasigroup.

# 8 Miscellaneous structures

## 8.1 The Octonions

The *Octonion quasigroup* $O_{16}$ consists of 16 elements $\{\pm 1, \pm i, \pm j, \pm k, \pm E, \pm I, \pm J, \pm K\}$, with the following multiplication table:

| | $1$ | $i$ | $j$ | $k$ | $E$ | $I$ | $J$ | $K$ |
|---|---|---|---|---|---|---|---|---|
| $1$ | $1$ | $i$ | $j$ | $k$ | $E$ | $I$ | $J$ | $K$ |
| $i$ | $i$ | $-1$ | $k$ | $-j$ | $I$ | $-E$ | $-K$ | $J$ |
| $j$ | $j$ | $-k$ | $-1$ | $i$ | $J$ | $K$ | $-E$ | $-I$ |
| $k$ | $k$ | $j$ | $-i$ | $-1$ | $K$ | $-J$ | $I$ | $-E$ |
| $E$ | $E$ | $-I$ | $-J$ | $-K$ | $-1$ | $i$ | $j$ | $k$ |
| $I$ | $I$ | $E$ | $-K$ | $J$ | $-i$ | $-1$ | $-k$ | $j$ |
| $J$ | $J$ | $K$ | $E$ | $-I$ | $-j$ | $k$ | $-1$ | $-i$ |
| $K$ | $K$ | $-J$ | $I$ | $E$ | $-k$ | $-j$ | $i$ | $-1$ |

Just as the Quaternions are commutative up to a sign, the Octonions are associative up to a sign. If we define the *associator* $\{a, b, c\} = \pm 1$ such that $a(bc) = \{a, b, c\}(ab)c$, then $\{a, b, c\} = a^2 b^2 c^2 (ab)^2 (bc)^2 (ca)^2 (abc)^2$ with the following properties:

1) It is invariant under permutations, so that $\{a, b, c\} = \{b, c, a\} = \{a, c, b\}$.

2) $\{a, a, b\} = 1$, so that $a(ab) = a^2 b$ and $a(ba) = (ab)a$. This makes $O_{16}$ an *alternative* algebra [12].

3) The associator of a product is the product of associators, i.e. $\{a, b, cd\} = \{a, b, c\}\{a, b, d\}$ (this was true for commutators in $Q_8$)

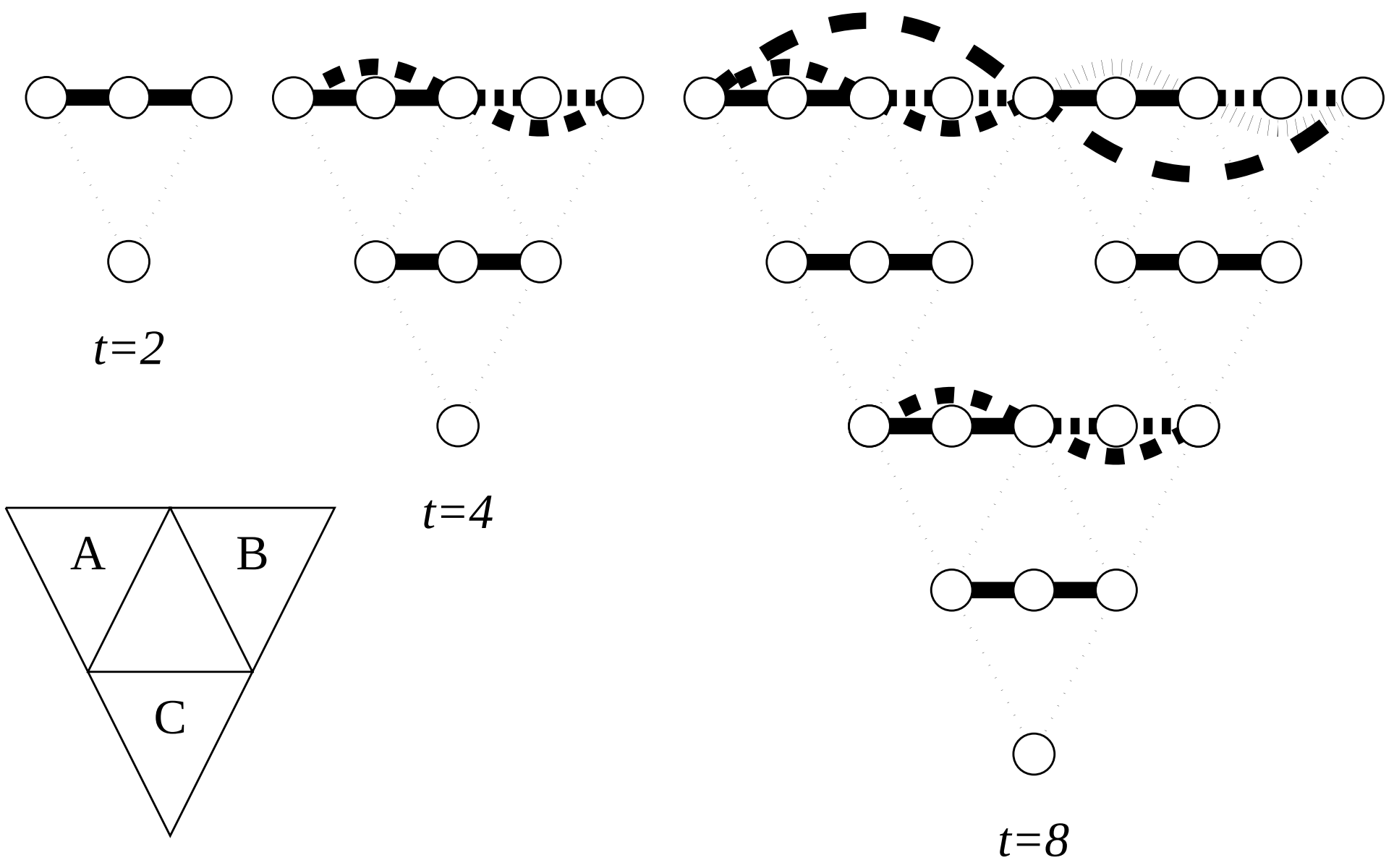

Figure 15: Scaling the set of associators for the Octonion CA. Each group of three elements corresponds to an associator.

4) $\pm 1$ associates and commutes with everything, and commutators, associators, and squares are all $\pm 1$.

Just as $Q_8/\{\pm 1\} = Z_2^2$, $O_{16}/\{\pm 1\} = Z_2^3$; so we can treat the Octonion CA as addition mod 2 on three components, with a prefactor of $\pm 1$.

For $t = 2$, we have

$$(a_0 a_1)(a_1 a_2) = \{a_0, a_1, a_1 a_2\} a_0 (a_1 (a_1 a_2)) = a_1^2 \{a_0, a_1, a_2\} \cdot a_0 a_2$$

since $\{a_0, a_1, a_1 a_2\} = \{a_0, a_1, a_1\}\{a_0, a_1, a_2\} = \{a_0, a_1, a_2\}$ and $a_1^2 = \pm 1$. For $t = 4$, the prefactor consists of the Quaternion prefactor $a_2^2 [a_1, a_3]$, times

$$\{a_0 a_2, a_1 a_3, a_2 a_4\}\{a_0, a_1, a_2\}\{a_2, a_3, a_4\}\{a_0, a_2, a_4\}$$

In general, the prefactor for $t = 2^{k+1}$ consists of $\{a_0, a_{2^k}, a_{2^{(k+1)}}\}$ times three copies of the prefactor for $t = 2^k$, with one copy ("A") as is, one ("B") where $a_i$ is replaced by $a_{i+2^k}$, and one ("C") where $a_i$ is replaced by $a_i a_{i+2^k}$. This is shown graphically in figure 15.

Note that the intermediate states we need for the associators in C are always products of intermediate states we needed for associators in A and B anyway. So each associator takes constant additional time to calculate in serial. Since the Quaternion part of the prefactor can be calculated as we did before, and the number of associators in the prefactor triples whenever we double $t$, we have

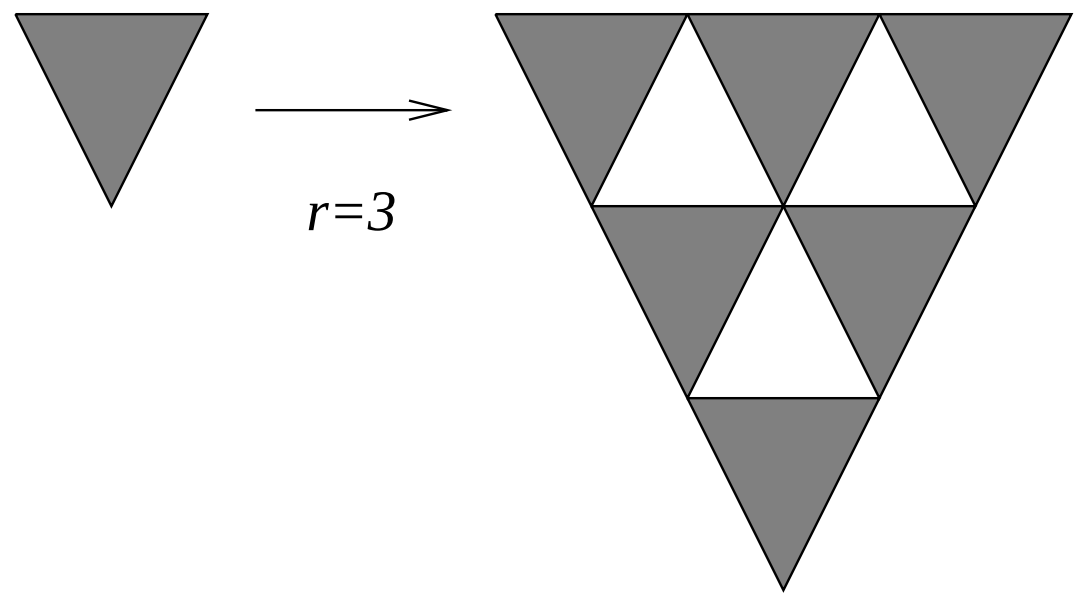

Figure 16: If, as for the Octonions, a CA scales except for a prefactor which associates and commutes with all its elements, then we conjecture multiplying $t$ by $r$ should give $r(r+1)/2$ copies of the prefactor expression. Here $r = 3$.

**Theorem 10.** *The Octonion CA can be predicted in serial time* $\mathcal{O}(t^{\log 3/\log 2})$.

In parallel, we can calculate the associators and intermediate states in A and B at the same time, and then calculate those in C in constant time; so doubling $t$ only adds a constant, and

**Theorem 11.** *The Octonion CA can be predicted in parallel time* $\mathcal{O}(\log t)$*, so its prediction is in* $\mathbf{NC}_1$.

These arguments would seem to hold whenever a CA is equivalent to a scaling CA except for commutators and associators which are in its *center* (the set of elements which commute and associate with everything). So we make the following conjecture:

**Conjecture.** *If a CA's commutators and associators are in its center, and if the CA mod its center scales with ratio r, then it can be predicted in serial time* $\mathcal{O}(t^\alpha)$ *where*

$$\alpha = \frac{\log r(r+1)/2}{\log r} < 2$$

*and parallel time* $\mathcal{O}(\log t)$*, so its prediction is in* $\mathbf{NC}_1$.

This value of $\alpha$ is based on figure 16, where we assume that $r(r+1)/2$ copies of the $t = r^k$ expression will be required in the $t = r^{k+1}$ expression.

(Since the associator of products is a product of associators, we could consider breaking down these expressions to a product of associators of single initial sites. In fact, this is how we got an algorithm for the Quaternions that runs in time $t \log^2 t$ instead of $t^{\log 3/\log 2}$. Unfortunately, in the case of the Octonions this seems to give a number of associators that grows by a factor of 6 each time we double $t$, giving $t^{\log 6/\log 2} > t^2$.)

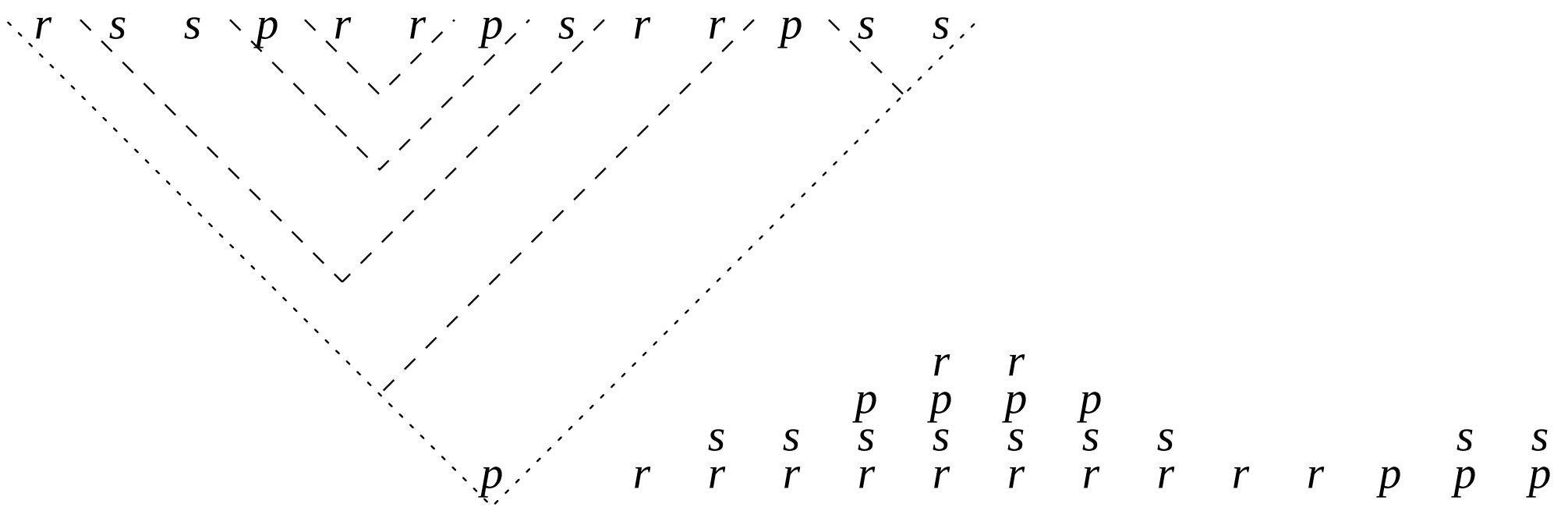

Figure 17: The rocks-scissors-paper CA, and the push-down automaton algorithm that predicts it. The stack represents successive right diagonals of the light-cone as we move through the input.

## 8.2 Idempotent semigroups

If $\bullet$ is associative and $aa = a$ for all $a$, then we have

$$\begin{array}{ccccccc} a_0 & & a_1 & & a_2 & & a_3 \\ & a_0 a_1 & & a_1 a_2 & & a_2 a_3 & \\ & & a_0 a_1^2 a_2 = a_0 a_1 a_2 & & a_1 a_2^2 a_3 = a_1 a_2 a_3 & & \\ & & & a_0 (a_1 a_2)^2 a_3 = a_0 a_1 a_2 a_3 & & & \end{array}$$

so $P_t = \prod_{i=0}^t a_i$.

Even without associativity, if the weaker property $(ab)(bc) = (ab)c$ holds, a simple induction shows that $P_t = \prod_i^{\text{right}} a_i = (\cdots((a_0 a_1) a_2) \cdots a_t)$, and similarly if $(ab)(bc) = a(bc)$ with the products on the left.

## 8.3 CAs with left- and right-moving domain walls

Suppose, for all $a$ and $b$, we have $ab = a$ or $ab = b$. Then every boundary between a domain of $a$s and $b$s moves left or right with a velocity 1/2. We say *$a$ dominates $b$ from the left* or $a \succ b$ if $ab = a$, and so on.

Such a system need not be associative or commutative. It is associative if $\prec$ and $\succ$ are transitive, and commutative if $a \succ b$ implies $b \prec a$. Figure 17 shows an example, the "rocks-scissors-paper" CA (see [9]), $r \succ s \succ p \succ r$, which is commutative but not associative: $r(sp) = rs = r$, but $(rs)p = rp = p$.

These CAs can be predicted in linear time in the following way, using a stack as memory. Start with an empty stack, and add $a_0$. Then read the $a_i$ from left to right: for each $a_i$, pop all the $a$s off the stack for which $a \prec a_i$, until we get to one for which $a \succ a_i$. Then add $a_i$ to the stack on top of it. Finally, the result $s$ is the symbol at the bottom of the stack.

Figure 16 shows how the algorithm works: the stack at each stage represents the rightmost diagonal of the light-cone bounded by $a_0$ and $a_i$. Since the total amount of pushing and popping is bounded by $t$, it just takes linear time.

Since this algorithm can be carried out by a *deterministic push-down automaton* (the finite state control can keep track of the bottom stack symbol) we have shown that, for each $s$, the set $L_s$ of initial sequences $w = a_0 a_1 \ldots a_t$ such that $P_t(w) = s$ is *deterministic context-free language* [15]. It can be generated by the following grammar, from the initial symbol $I_s$ and symbols $s, s', X_{s\succ}, X_{\prec s}$, and $X_{s\succ,\prec t}$ for all CA states $s$ and $t$:

$$I_s \to X_{\prec s} s' X_{s\succ}$$

$$X_{\prec s} \to X_{\prec t} t' \text{ or } t' X_{t\succ,\prec s} \text{ if } t \prec s$$

$$X_{s\succ} \to X_{s\succ,\prec t} t' \text{ or } t' X_{t\succ} \text{ if } s \succ t$$

$$X_{s\succ,\prec t} \to X_{s\succ,\prec u} u' \text{ or } u' X_{u\succ,\prec t} \text{ if } s \succ u \prec t$$

Here the $X$s represent places where we can add new domains, tracing backwards from the final site, that are annihilated (from the left, the right, or both) by the domains we already know about. To these productions, we add $X \to \epsilon$ (the empty string) to erase the $X$s at the end, and $s' \to ss'$ or $\epsilon$ to create multiple sites in each domain (with only unprimed symbols in the final product).

This grammar is *unambiguous*, meaning that each sequence $w$ has a unique derivation tree using these productions. Unambiguous context-free languages can be recognized in $\mathcal{O}(\log |w|)$ parallel time [18]; in addition, this PDA is *input-driven*, meaning that the change in its stack height depends only on the current input symbol, and for such PDAs this can be done with $\mathcal{O}(|w|)$ processors [18]. Moreover, the above algorithm clearly takes linear time on a two-tape Turing machine. So

*Theorem 12. CAs with the property that $ab = a$ or $b$, so that domain boundaries move at constant speed and annihilate, can be predicted in time $\mathcal{O}(t)$ in serial or $\mathcal{O}(\log t)$ in parallel (with $\mathcal{O}(t)$ processors), so their prediction is in* $\mathbf{NC}_1$.

Since a two-dimensional version of the rocks-scissors-paper CA generates spiral waves, it's interesting to ask whether spiral waves in general, or any CA model of them, can be efficiently predicted or parallelized: are they, for instance, in **NC**?

# 9 Scaling relations and the medial identity

As stated briefly above, a CA *scales with ratio* $r$ if $P_r(a_0, \ldots, a_r) = a_0 \bullet a_r$, so that $r$ steps is equivalent to one step on the leftmost and rightmost initial conditions. As we saw in section 3, scaling allows us to predict a CA in linear time.

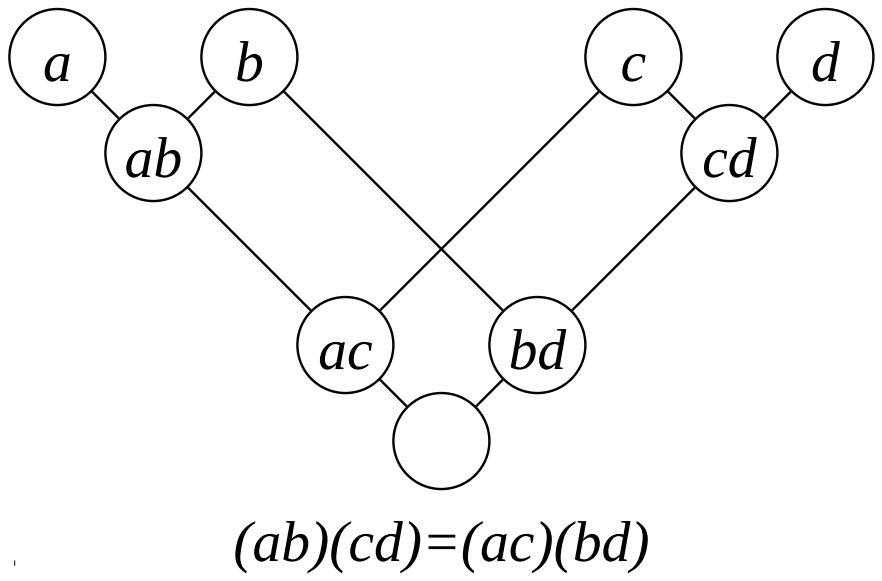

Figure 18: The medial identity follows from scaling. To predict $r+1$ time-steps, we can use the scaling relation on the first $r$ or the last $r$, and these must yield the same result.

What algebraic properties must a scaling CA have? Consider $P_{r+1}(a, b, \ldots, c, d)$. As figure 18 shows, we can use the scaling either for the first $r$ steps, or the last $r$. Both these methods must yield the same result, so the *medial identity* holds

$$(a \bullet b) \bullet (c \bullet d) = (a \bullet c) \bullet (b \bullet d)$$

which we ran across earlier in our discussion of Steiner systems. (Mediality alone can't help us predict a CA, since all the products in (1) are of the form $(ab)(bc)$.)

Any medial quasigroup with identity is an Abelian group, since we immediately get associativity and commutativity: $a(bc) = (ae)(bc) = (ab)(ec) = (ab)c$, and $ab = (ea)(be) = (eb)(ae) = ba$. In addition, any quasigroup separably isotopic to an Abelian group, i.e. of the form $a \bullet b = f(a) + g(b) + h$ where $f$ and $g$ are homomorphisms and $h$ is a constant, is medial if $f$ and $g$ commute.

In fact, it can be shown that all medial quasigroups are of this form [12], with $h = 0$ if there is an idempotent element $ee = e$ (which becomes the identity of the group). So the only quasigroup CAs which have a scaling relation are those we already learned how to deal with in section 5.1. However, there could be non-quasigroup CAs which scale (i.e., medial groupoids) for which scaling gives the only fast algorithm.

In any case, we might as well state what we've already proved:

**Theorem 13.** *Any CA with a scaling relation satisfies the medial identity. Furthermore, it can be predicted in $\mathcal{O}(t)$ serial time or $\mathcal{O}(\log t)$ parallel time, and so its prediction is in* $\mathbf{NC}_1$.

# 10 Principles of superposition

Just as for scaling, we can ask what algebraic properties are required for a CA to have a principle of superposition.

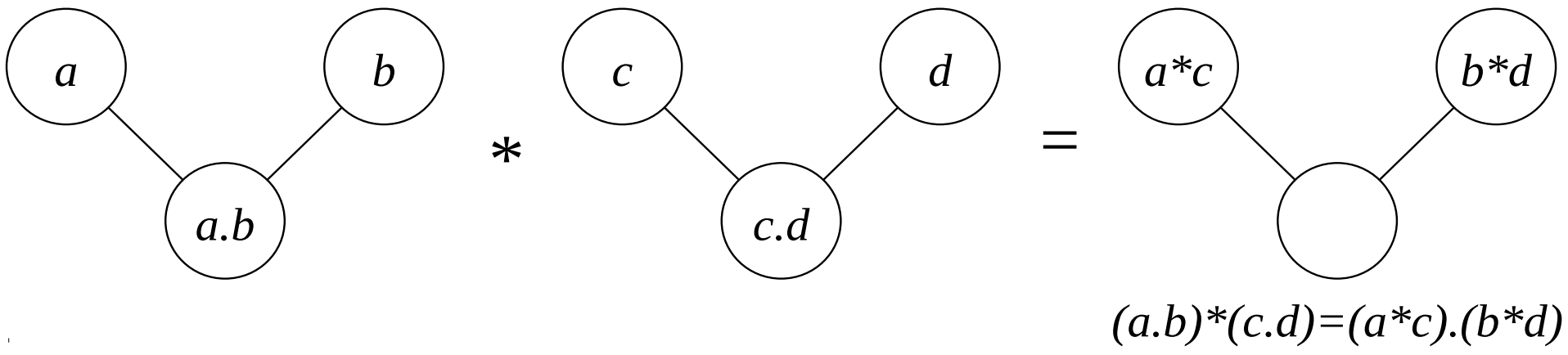

Figure 19: The generalized medial identity (9) follows from a principle of superposition. We can apply the superposition operation $*$ before or after the CA rule $\bullet$.

In general, a principle of superposition means that the state space has some structure on which the CA rule, seen as a function $f : A^2 \to A$ (or on sequences $A^n \to A^{n-1}$, or on bi-infinite suquences $A^Z \to A^Z$) is a homomorphism. So, for instance, if there is a binary operation $*$ on the alphabet $A$, we write $f((a, b) * (c, d)) = f(a, b) * f(c, d)$. Writing $a \bullet b$ for $f(a, b)$, we get a pleasantly symmetric identity

$$(a \bullet b) * (c \bullet d) = (a * c) \bullet (b * d) \tag{9}$$

as shown in figure 19 — a kind of generalized medial identity with two operations. If $\bullet$ and $*$ are the same the normal medial identity holds, as was pointed out in [10]. (Incidentally, (9) also implies that $*$ and $\bullet$ commute as CAs, since $(a \bullet b) * (b \bullet c) = (a * b) \bullet (b * c)$.)

This identity (9) alone doesn't tell us very much. However, there are a few assumptions we can reasonably make.

First of all, in order for a principle of superposition to be useful, we need a notion of a *$\delta$-function* $\delta_{x,a}$: namely, an initial condition which is "zero" (an identity) except at site $x$, where it is $a$. Then any initial condition can be built up out of the $\delta$s, $\delta_{1,a_1} * \delta_{2,a_2} * \cdots * \delta_{t,a_t} = (a_1, a_2, \ldots, a_t)$.

So $*$ must have an identity, $e_*$, so that $\delta_{x,a}$ can be $e_*$ everywhere but at $x$. But then

$$a \bullet b = (a * e_*) \bullet (e_* * b) = (a \bullet e_*) * (e_* \bullet b)$$

Moreover,

$$e_* \bullet (a * b) = (e_* * e_*) \bullet (a * b) = (e \bullet a) * (e \bullet b)$$

So the function $f(a) = a \bullet e_*$ is a homomorphism with respect to $*$, and similarly for $g(a) = e_* \bullet a$. But this is just

$$a \bullet b = f(a) * g(b)$$

again — in other words, $\bullet$ is homomorphically isotopic to $*$.

Counterintuitively, $*$ is not necessarily associative or commutative. For instance, if $a \bullet b = h(b)$, then (9) holds whenever $h$ is a homomorphism of $*$.

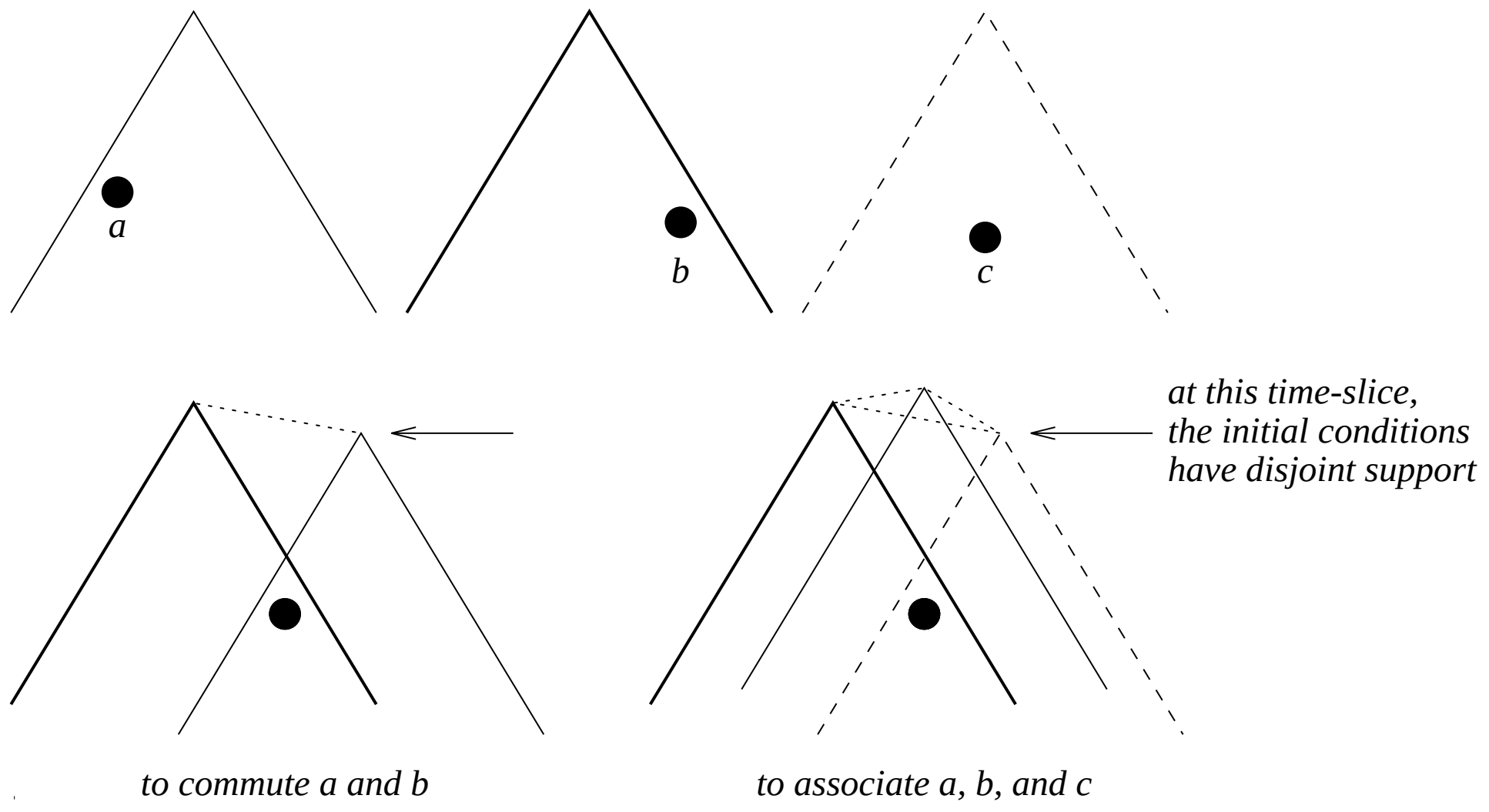

Figure 20: Overlaying initial conditions with disjoint support to get commutativity or associativity on elements whose positions in the Green's function are separated by spacelike intervals (the dotted lines).

However, we can show that $*$ commutes and associates on some elements: for instance, we have

$$(a \bullet e_*) * (e_* \bullet b) = (a * e_*) \bullet (e_* * b) = (e_* * a) \bullet (b * e_*) = (e_* \bullet b) * (a \bullet e_*)$$

so $*$ commutes on pairs of the form $a \bullet e_*$, $e_* \bullet b$. This symbolizes the fact that it doesn't matter what order we $*$ together the $\delta$ functions $(a, e_*)$ and $(e_*, b)$; since each one is $e_*$ where the other isn't, and $e_*$ commutes with everything. Similarly, $*$ must be associative on any three elements from the $t = 2$ row of the Green's function, of the form $(e_*e_*)(e_*a)$, $(e_*b)(be_*)$, and $(ce_*)(e_*e_*)$.

More generally, $*$ is commutative on any two $\delta$s with different $x$, and associative on any three; and since we can go down to any row where their Green's functions overlap, we know that $*$ is commutative on any two elements that occur in the same row of the Green's functions of some two elements, and associative on any three.

We can generalize this further to pairs or triplets that appear in Green's functions at positions with *spacelike intervals* between them; i.e., $\Delta x > \Delta t$ or $\Delta x < 0$ if $\Delta t \geq 0$, and $\Delta x < \Delta t$ or $\Delta x > 0$ if $\Delta t < 0$. As shown in figure 20, we can overlay two or three initial conditions with disjoint support (each is $e_*$ wherever the others aren't) to put these elements in the same place; since $*$ associates and commutes on these initial conditions, it must also on these elements.

Since any spreading of the Green's function will lead to a spacelike interval,

it appears that $*$ will be associative and commutative except on parts of the Green's function that don't spread out at all; for instance, with $a \bullet b = h(b)$, where information is only travelling to the left (or with an $r = 1$ CA we could construct an example with a non-commuting part travelling straight down). This kind of behavior doesn't add much interest, and is certainly no problem to predict, being just an identity map on sequences with each symbol being mapped by $h$.

Finally, if we know $*$ to be a quasigroup, then associativity and commutativity make it an Abelian group; then $\bullet$ is homomorphically isotopic to an Abelian group, and the general method of theorem 6 can be used to predict it.

A few other lemmas of some interest, assuming a principle of superposition (9) and an identity $e_*$, are:

**Lemma 14.** *If $\bullet$ is a quasigroup, $*$ is an Abelian group.*

**Proof.** If $\bullet$ is a quasigroup, then $\bullet(a, b) = a \bullet b$ is one-to-one on both its inputs. But $a \bullet b = (a \bullet e_*) * (e_* \bullet b)$ and $a \bullet e_*$ and $e_* \bullet b$ can take any values $c$ and $d$, so $c * d$ must be one-to-one on both $c$ and $d$. So $*$ is a quasigroup with identity, or a loop; moreover, both $\bullet$ and $*$ are isotopic to an Abelian group (by a theorem of Sade [13] following from (9)) and loops that are isotopic to groups are isomorphic to them [12]. So $*$ is an Abelian group.

**Lemma 15.** *If every element $a$ can be expressed as a product $b \bullet c$, then $e_*$ is idempotent under $\bullet$ (i.e., $e_* \bullet e_* = e_*$).*

**Proof.** $(e_* \bullet e_*) * a = (e_* \bullet e_*) * (b \bullet c) = (e_* * b) \bullet (e_* * c) = b \bullet c = a$. So $e_* \bullet e_*$ is also an identity of $*$, but identities are unique.

An element that can't be expressed as a product $b \bullet c$ can only appear in the initial row. So we can take these elements out of $A$, move forward one step, and continue with $e_* \bullet e_*$ as our new identity on the remaining subset of $A$.

**Lemma 16.** *If $\bullet$ has a left (right) identity $e_\bullet$, then $e_*$ is also a left (right) identity of $\bullet$.*

**Proof.** Suppose $e_\bullet$ is a left identity of $\bullet$. Then $a = e_\bullet \bullet a = (e_\bullet \bullet e_*) * (e_* \bullet a) = e_* * (e_* \bullet a) = e_* \bullet a$, so $e_*$ is also. Similarly for right identities.

**Corollary.** *If $\bullet$ has an identity, then $e_\bullet = e_*$ and $\bullet$ and $*$ are identical.*

**Proof.** Since $e_\bullet$ is both a left and a right identity of $\bullet$, so is $e_*$. So $e_\bullet \bullet e_* = e_\bullet = e_*$ so they're the same $e$. Then $a \bullet b = (a \bullet e) * (e \bullet b) = a * b$.

# 11 Conclusion

We have derived the following computation times for CAs with various properties:

| CA property | serial | parallel |
|---|---|---|
| Left- and right-moving domain walls | $t$ | $\log t$ |
| Scaling | $t$ | $\log t$ |
| Abelian group or semigroup | $t$ | $\log t$ |
| Isotopic to an Abelian group: | | |
| with a commuting row | $t$ | $\log t$ |
| logarithmic scaling | $t \log t$ | $\log t \, \log\log t$ |
| in general | $\begin{cases} t \log t \log\log t \text{ (TM)} \\ t \log t \text{ (RAM)} \\ t \text{ (SMM)} \end{cases}$ | $\log^2 t$ |
| Quaternions | $t \log^2 t$ | $\log t$ |
| Octonions | $t^{\log 3/\log 2}$ | $\log t$ |
| Direct simulation | $t^2$ | $t$ |

We have shown that some nonlinear rules, like the Quaternion CA or CAs with moving domain walls, are in fact easily predictable. We have also shown that linearity alone does not assure linear-time predictability; CAs with complex Green's functions can take time up to $t \log t \log\log t$ on a Turing machine or $\log^2 t$ in parallel. In addition, in the parallel case we can make some distinctions based on the number of processors; for instance, $\mathcal{O}(t)$ for Abelian and domain-wall rules, and $\mathcal{O}(t^2)$ for vector-valued CAs and the Quaternions. Finally, we have explored the algebraic consequences of macroscopic properties like scaling and superposition.

Is it possible that all CA rules can be predicted more efficiently than by direct simulation? In particular, that all CA rules are in **NC**, meaning their parallel computation time is $\log^j t$ for some $j$? The precise answer is *only if* **P** = **NC**, i.e. only if all problems solvable in polynomial serial time can be efficiently parallelized (solved in polylogarithmic parallel time).

This is because the general problem of predicting a CA for $t$ time-steps is *P-complete*: any polynomial serial time problem can be reduced to it. This is easy to see with a CA that simulates a Turing machine, such as in [4]. Consider a Turing machine program that runs in time $n^k$ on input of length $n$. Pad the input with blanks so that the input size is $n^k$ and the time is equal to the input size. If this CA is in $\mathbf{NC}_j$ we'll get our answer in $k^j \log^j n$ parallel time, showing that the original problem is in $\mathbf{NC}_j$.

Another perspective is to think of CAs as Boolean circuits, with gates at each site in space-time. Calculating the value of such a circuit is P-complete; however, calculating the value of a Boolean expression is in $\mathbf{NC}_1$. So all CAs (and all problems in **P**) are in $\mathbf{NC}_1$ if and only if circuits of size $n$ can always be written as expressions of size $n$. Although circuits of size $n$ would *a priori* become expressions of size $2^n$ (since we have to repeat a sub-expression every time the circuit accesses the corresponding gate) just as the expressions in (1) grow exponentially with $t$, it hasn't actually been proven that this is necessary. Whether **NC** = **P** remains as open, and as important, an unsolved problem of complexity theory as whether $\mathbf{P} = \mathbf{NP}$.

In the meantime, we pose several open questions.

1) We conjecture that CAs based on any finite group are in $\mathbf{NC}_k$ for some $k$. We show in [19] that CAs based on nilpotent and solvable groups are in $\mathbf{NC}_1$ and $\mathbf{NC}_2$ respectively, but it is still unclear whether all finite groups are efficiently predictable.

2) In analogy with the rocks-scissors-paper CA, spiral waves in two dimensions seem to be intermediate in complexity between linearity and computational universality. How hard are they to predict?

3) By grouping pairs of sites together, we can turn the two-state, $r = 1$ elementary CA rules of [11] into algebras with four elements. For which of these 256 rules do these algebras have enough structure to be predicted efficiently? The algebraic relationship of these "block maps" and the underlying CA is explored in [20].

We hope through this kind of approach to develop an understanding of what kinds of nonlinear dynamics have enough structure to be easily understood and predicted, and to create a hierarchy of nonlinear systems of increasing complexity in the same spirit as the space and time hierarchies of algorithmic complexity theory [16] and the Chomsky hierarchy of language and automata theory [15].

**Acknowledgements.** I am grateful to Mats Nordahl for insightful discussions, especially on the Octonions section, and to Elizabeth Hunke and Spootie for inspiration and comfort.